\documentclass[oneside,12pt]{article}
\usepackage[utf8]{inputenc}
\usepackage[top=1in,bottom=1in,left=1in,right=1in]{geometry}
\usepackage{authblk}
\usepackage{natbib} 

\usepackage{placeins}
\usepackage{graphicx,comment}
\usepackage{caption}
\usepackage{subcaption}
\usepackage{float}

\usepackage{amsfonts, amsmath, amssymb, amsthm, constants, bbm}
\usepackage{mathrsfs}
\usepackage{enumitem}
\usepackage{booktabs}

\setlist[itemize]{topsep=0mm}
\setlist[enumerate]{topsep=0mm}

\usepackage{xcolor}
\definecolor{darkred}{RGB}{100,0,0}
\definecolor{darkgreen}{RGB}{0,100,0}
\definecolor{darkblue}{RGB}{13,53,217}

\usepackage{hyperref}
\hypersetup{colorlinks=true, linkcolor=darkred, citecolor=darkgreen, urlcolor=darkblue}
\usepackage{xurl}

\newcommand\numberthis{\addtocounter{equation}{1}\tag{\theequation}}

\newtheorem{thm}{Theorem}

\newtheorem{lem}{Lemma}

\newtheorem{defi}{Definition}

\theoremstyle{remark}
\newtheorem{rem}{Remark}

\def\cD{\mathcal{D}}

\def\cN{\mathcal{N}}

\def\cX{\mathcal{X}}

\def\bX{\mathbf{X}}
\def\bY{\mathbf{Y}}

\def\bx{\mathbf{x}}

\renewcommand{\P}[1]{\mathbb{P}\left(#1\right)}
\newcommand{\ind}[1]{\mathbbm{1}\left\{ #1 \right\}}
\newcommand{\given}{\;\middle|\;}
\newcommand{\givenB}[1]{\;#1\vert\;}

\usepackage{mathtools}
\DeclarePairedDelimiter\abs{\lvert}{\rvert}%
\DeclarePairedDelimiter\norm{\lVert}{\rVert}%
\makeatletter
\let\oldabs\abs
\def\abs{\@ifstar{\oldabs}{\oldabs*}}
\let\oldnorm\norm
\def\norm{\@ifstar{\oldnorm}{\oldnorm*}}
\makeatother

\newcommand{\defeq}{\vcentcolon =}

\usepackage[T1]{fontenc}
\usepackage{lmodern}
\makeatletter
\newcommand{\twobar}{/\kern-0.2em/} 
\let\orig@Url@acthash\Url@acthash 
\let\new@Url@acthash\Url@acthash 
\g@addto@macro{\new@Url@acthash}{\Url@Edit\Url@String{//}{\twobar}}
\let\orig@urlstyle\urlstyle 
\def\urlstyle{\let\Url@acthash\orig@Url@acthash\orig@urlstyle}
\g@addto@macro{\url@rmstyle}{\let\Url@acthash\new@Url@acthash} 
\g@addto@macro{\url@sfstyle}{\let\Url@acthash\new@Url@acthash}
\makeatother

\begin{document}

\title{Inference with Sequential Monte-Carlo Computation of $p$-values: Fast and Valid Approaches}
\author[1]{Ivo V. Stoepker}
\author[1]{Rui M. Castro}
\affil[1]{\small Department of Mathematics and Computer Science, Technische Universiteit Eindhoven, Eindhoven, The Netherlands} 
\date{}
\maketitle

\thispagestyle{empty}

\vspace{-1cm}

\begin{abstract}
\vspace{-0.2cm}
Hypothesis tests calibrated by (re)sampling methods (such as permutation, rank and bootstrap tests) are useful tools for statistical analysis, at the computational cost of requiring Monte-Carlo sampling for calibration. It is common and almost universal practice to execute such tests with predetermined and large number of Monte-Carlo samples, and disregard any randomness from this sampling at the time of drawing and reporting inference. At best, this approach leads to computational inefficiency, and at worst to invalid inference. That being said, a number of approaches in the literature have been proposed to adaptively guide analysts in choosing the number of Monte-Carlo samples, by sequentially deciding when to stop collecting samples and draw inference. These works introduce varying competing notions of what constitutes ``valid'' inference, complicating the landscape for analysts seeking suitable methodology. Furthermore, the majority of these approaches solely guarantee a meaningful estimate of the testing outcome, not the $p$-value itself --- which is insufficient for many practical applications. In this paper, we survey the relevant literature, and build bridges between the scattered validity notions, highlighting some of their complementary roles. We also introduce a new practical methodology that provides an estimate of the $p$-value of the Monte-Carlo test, endowed with practically relevant validity guarantees. Moreover, our methodology is sequential, updating the $p$-value estimate after each new Monte-Carlo sample has been drawn, while retaining important validity guarantees regardless of the selected stopping time. We conclude this paper with a set of recommendations for the practitioner, both in terms of selection of methodology and manner of reporting results.

\vspace{2mm}\noindent\textbf{Keywords:} Monte-Carlo Calibration, Sequential Testing, Confidence Sequence, Resampling Risk
\end{abstract}

\vspace{-0.7cm}
\section{Introduction}

The vast majority of statistical analysis involves testing procedures. Results are generally reported by means of a $p$-value quantifying the incompatibility of observed outcomes with a set hypothesis. In many cases, test calibration requires accurate distributional knowledge of the corresponding test statistics (under the null hypothesis). More often than not, even under relatively benign models and settings, the characterization of such distributions relies on asymptotic considerations, and more modernly on the use of computational tools. This paper focuses on the latter scenarios, and specifically on the use of Monte-Carlo sampling for test calibration. To use such tests, the analyst needs to draw samples from a (possibly complex) random process for calibration, which is often computationally demanding. A natural question to ask is how should one choose the number of random samples, and given that choice (which can be data-dependent) how to ensure the ensuing testing results are reported properly. We proceed with a motivating example in Section~\ref{sec:motex} and the general setting is introduced in Section~\ref{sec:setting}.

It is common and almost universal practice to execute such tests with predetermined and large number of Monte-Carlo samples, and disregard any randomness from this sampling at the time of drawing and reporting inference. At best, the downside of this approach is mostly computational inefficiency. More troublesome, however, is that commonly used samples sizes may lead to invalid inference. We illustrate this in Section~\ref{sec:fixed-sample}.

Previous authors have tried to alleviate this issue, and as a result multiple approaches in the literature exist that aim to meaningfully guide analysts in choosing the number of Monte-Carlo samples, while still producing valid inference. Such methods sequentially draw Monte-Carlo samples until reliable inference can be made. This can allow for significant computational savings. Despite their appeal, these works introduce varying (and often competing) notions of what constitutes ``valid'' inference. Unfortunately, the up-to-now disconnected landscape complicates the selection of a suitable methodology by the analyst.

Furthermore, the majority of these methods solely guarantee a meaningful estimate of the binary accept/reject testing outcome, which is insufficient in many practical scenarios \citep{Wasserstein2016, Greenland2016, Aguinis2021}. 
Possibly to account for this, implementations of such methods typically do supply a $p$-value estimate --- but this estimate often carries little information and is unstable under replication of the analysis. This may deter analysts from using such methods altogether. We illustrate this in Section~\ref{sec:goals}.

\textbf{Contributions:} We contribute solutions aiming to solve the problems highlighted above:
\begin{enumerate}
\item We connect the disconnected landscape in the literature by comparing, contrasting and extending the varying competing notions of ``valid'' inference previously proposed by other authors. We conclude this survey with a set of recommendations for the practitioner, both in terms of how methodology should be selected, as well as how results should be reported.
\item We propose a new simple and practical methodology providing an estimate of the $p$-value of the Monte-Carlo test, endowed with practically relevant validity guarantees. The proposed methodology is sequential by updating its $p$-value estimate after each new Monte-Carlo sample has been drawn. It remains valid regardless of how the number of Monte-Carlo samples are chosen; in particular, it may be decided sequentially based on the values of the estimates seen thus far.
\end{enumerate}

The remainder of this section is devoted to further illustrating the setting we study, and the associated problems as sketched above.

\subsection{A motivating example: scan statistics}\label{sec:motex}
To set the stage, we start with a motivating example before introducing a general framework. This example is meant for illustration purposes, and it crudely oversimplifies important aspects for the sake of clarity.

Suppose we want to monitor the incidence of a certain type of crime in a country, and determine if it is abnormally high in certain regions. Our region of interest consists of subregions indexed by $K = \{1,\dots k\}$ and we have recorded the number of events over some period of time in these subregions by $\bX = \{X_1,\dots,X_k\}$. A natural starting point would be to test the null hypothesis that these observations arose from a homogeneous\footnote{For the crime incidence example this is an oversimplification, that does not take into account the spatially varying population density in the subregions, for instance.} Poisson process with mean $\mu_0$. For the sake of simplicity we assume $\mu_0$ is known in this example. Therefore, the null hypothesis is
\[
H_0: \forall_{i\in K} : X_i \overset{\text{i.i.d.}}{\sim} \text{Poisson}(\mu_0) \ .
\]
Our context suggests we wish to be powerful against spatial clusters in the distribution. A powerful way to proceed in this situation is through the use of scan statistics \citep{Kulldorff1997, Kulldorff1999, Abolhassani2021}. For clarity of presentation we consider the simple\footnote{In \cite{AriasCastro2013} it is argued this simple version is in some sense asymptotically equivalent to the likelihood-ratio based scan statistic introduced by \cite{Kulldorff1997}.} scan statistic 
\begin{equation}\label{eq:scan}
S_\cD(\bX) \defeq \max_{A \in \cD} \frac{1}{\sqrt{\abs{A}}}\sum_{i\in A}\big( X_i - \mu_0 \big) \ ,
\end{equation}
where $\cD$ denotes a set of subsets of the studied subregions $K$ containing all candidate crime clusters we would wish to investigate. It is generally not desirable to investigate the existence of clusters over all possible subsets of $K$, but rather over subsets restricted by size and shape (for example size-constrained contiguous subsets of $K$ with respect to the geometry of the subregions studied). This complicates the computation of the maximum in~\eqref{eq:scan}.

Large values of $S_\cD(\bX)$ are incompatible with the homogeneity assumption. Given our simple null hypothesis, a suitable $p$-value for observed data $\bX$ can be computed as
\[
p(\bX) \defeq \mathbb{P}_{\bY \sim H_0}\bigg( S_\cD(\bY) \geq S_\cD(\bX) \givenB{\bigg} \bX \bigg) \ ,
\]
where $\bY$ is independent from $\bX$. While our null hypothesis for the data is uncomplicated, a closed-form finite sample expression for the distribution of the scan statistic $S_\cD(\bY)$ for~$\bY$ sampled under the null hypothesis is unknown, precluding analytical computation of the probability above. However, it is straightforward (but computationally demanding) to calibrate the test via Monte-Carlo sampling. Specifically, one can (repeatedly) sample from the distribution of $S_\cD(\bY)$ under the null hypothesis through first sampling null data observations $\bY_i \sim H_0$ and subsequently compute the scan statistics on the simulated datasets $S_\cD(\bY_i)$. The $p$-value above can be approximated\footnote{There are good reasons \citep{Phipson2010, North2002} to modify $p$-value estimate in~\eqref{eq:p-est-intro-simple} by adding 1 to both the numerator and denominator, but for simplicity (and since the problems we illustrate do not depend on it) we are using a naive estimator here. In Section~\ref{sec:fixed-predet} we discuss these estimators in more detail.} by a Monte-Carlo estimator
\begin{equation}\label{eq:p-est-intro-simple}
\hat p(\bX) = \frac{\sum_{i=1}^N \ind{ S_\cD(\bY_i) \geq S_\cD(\bX) }}{N} \ .
\end{equation}
The computational load in this setting can be considerable, which comes primarily from the (repeated) computation of the test statistic $S_\cD$, involving a costly maximization (which is typically impossible to execute greedily or otherwise efficiently) over a potentially large set of subsets $\cD$.

Related literature typically uses $N=1000$ Monte-Carlo samples for illustration purposes: see, for example, \cite{Edgington2020, Ernst2004, Phipson2010, Abolhassani2021, Kulldorff1995, Kulldorff1999, Wilcox2009, Besag1989, Good2000, Marriott1979} where methods are illustrated with sample sizes of this order of magnitude, or \cite{Marozzi2004} and references therein for such recommendations. However, as we will illustrate in Section~\ref{sec:fixed-sample}, interpreting these illustrations as general recommendations for practice may not be suitable, and it is possibly misleading. Nevertheless, at this point we may proceed with the intent of sampling 1000 Monte-Carlo samples and deciding to test at a $5\%$ significance level. Now, suppose that after 30 null samples have been drawn, none of these 30 Monte-Carlo samples $S_\cD(\bY_i)$ exceed the value of the scan statistic for the original data $S_\cD(\bX)$. Given the independence between samples, one may start to feel quite confident that if one would proceed and draw the remaining 970 samples, the eventual $p$-value estimate will not exceed our target significance level of $5\%$. It may be tempting to stop the computational efforts at this point --- but does our inference remain valid? And, if we stop (either prematurely or after 1000 samples), what outcome should we report from this test? We aim to provide answers to these questions in the remainder of this work, with an eye for simple and meaningful guidelines for analysts and practitioners.

This example is prototypical of the type of Monte-Carlo tests for which determining the number of samples to use is most pertinent. In particular, the high computational effort required to compute the test statistic on the generated samples is what might make the choice for the number of Monte-Carlo samples practically relevant. This is only exacerbated when more intricate statistics are used instead. For example, \cite{Tango2005} remark computations for their scan statistic can take up to a week when large potential cluster sizes are considered. Computationally demanding test statistics arise in other contexts as well; for example, goodness-of-fit tests calibrated by the bootstrap \citep{Stute1993}, as well as tests calibrated by permutations \citep{AriasCastro2018, Stoepker2023, Stoepker2024}.

\subsection{Calibration by Monte-Carlo simulation}\label{sec:setting}

We now proceed to introduce our framework in full generality. This abstraction illustrates the ubiquity of settings where calibration by Monte-Carlo simulation is leveraged and pinpoints the common factor across all such settings. We introduce notation we use throughout our remaining discussion.

Suppose we are able to collect data $\bX$ generated from some unknown stochastic process $F\in\Omega$. Let $\mathcal{X}$ denote the set of possible values data can take (i.e. $\mathcal{X}\defeq\bigcup_{F\in \Omega} \text{support}(\bX)$). We are interested in testing the (null) hypothesis that $F$ belongs to a subclass class of processes $\Omega_0\subset \Omega$:
\[
H_0: F \in \Omega_0 \ .
\]
To test the above hypothesis, we collect data on which we base the computation of a test statistic denoted by $T(\bX)$. We consider a testing scenario where we reject the above null hypothesis for large values of $T(\bX)$. If the above null hypothesis is composite, one could proceed through a conservative worst-case calibration\footnote{This amounts to computing a $p$-value as 
$$
p(\bX) = \sup_{F \in \Omega_0} \mathbb{P}_{\bY \sim F} \Big(T(\bY) \geq T(\bX) \hspace{0.2cm}\Big\vert\hspace{0.2cm} \bX \Big) \ .
$$
See, for example, Theorem 10.12 in \cite{wasserman2010}. Inference based on such worst-case calibration may be (highly) conservative and is typically only attempted when $\Omega_0$ is of some parametric form \citep{Berger1994,Bayarri2000}.
}, but a more effective way forward is by finding a distribution $F_0(\bX)$, possibly dependent on our observations $\bX$, such that under \emph{all} models under the null hypothesis $F\in\Omega_0$ the following quantity stochastically dominates the continuous uniform distribution on the unit interval\footnote{This implies for all $F\in\Omega_0$ and $t\in[0,1]: \P{p^*(\bX) \leq t} \leq t$. Ideally the $p$-value is uniformly distributed under the null, leading to an exact test. Stochastic domination of the uniform distribution is more general and implies the ensuing $p$-value can be exact or conservative.}:
\begin{equation}\label{eq:p_val_true}
p^*(\bX) \defeq  \mathbb{P}_{\bY \sim F_0(\bX)}\bigg( T(\bY) \geq T(\bX) \hspace{0.2cm}\Big\vert\hspace{0.2cm} \bX \bigg) \ .
\end{equation}
Though it may seem challenging to find a (conditional) distribution $F_0(\bX)$ such that the $p$-value above is valid --- particularly when stated in the abstract here --- finding such distributions is usually natural in a concrete context, and as a result this approach is common. For example, when the null models satisfy some exchangeability properties, a common choice of $F_0(\bX)$ is a permutation distribution on the observations $\bX$.

At this point, it is not important how to find such distributions $F_0(\bX)$, but rather to recognize how various tests, such as those based on calibration by permutation, are fundamentally based on a $p$-value of the form and interpretation of~\eqref{eq:p_val_true}, and we defer further comments regarding this to Remark~\ref{rem:exch}.

Next, given a significance level $\alpha\in(0,1)$, chosen before inspecting the data, we might also report the decision to accept or reject the null hypothesis through computation of
\begin{equation}\label{eq:phi}
\phi^*_{\alpha}(\bX) \defeq \ind{ p^*(\bX) \leq \alpha} \ ,
\end{equation}
where $\phi^*_\alpha(\bX) = 1$ implies the null hypothesis is rejected. We refer to $p^*(\bX)$ as the ``true'' $p$-value and to $\phi^*_{\alpha}(\bX)$ as the ``true'' accept/reject decision. Ideally the analyst would analytically compute and report these quantities, leading to conventional validity and interpretation of their inference. 

In the context we consider, the description of the distribution of $T(\bY)$ when $\bY \sim F_0(\bX)$ is elusive --- which precludes analytic computation of~\eqref{eq:p_val_true}
 --- but one has the ability to \emph{sample} from the distribution through sampling from $F_0(\bX)$ and subsequently computing $T(\bY)$. As such, while the $p$-value~\eqref{eq:p_val_true} and subsequent decision~\eqref{eq:phi} cannot be computed exactly, we can instead obtain samples from the Bernoulli distribution\footnote{In the context of sampling for the purposes of calibration by permutation the proposed Bernoulli samples may seem not to apply. It is true that for permutation tests, distinct permutations could be sampled instead to estimate~\eqref{eq:p_val_true}, which would then lead to samples from a hypergeometric distribution. However this is not a common practice as the effort required to sample distinct permutations is typically not worth the nominally faster convergence to the true $p$-value~\eqref{eq:p_val_true}.}
\begin{equation}\label{eq:mc_samples}
B_i|\bX \sim \text{Bernoulli}\Big(p^*(\bX)\Big) =\vcentcolon G^*(\bX) \ , 
\end{equation}
where $i\in\{1,2,\ldots\}$. It is precisely this (independent) sampling that defines such tests as ``calibrated by Monte-Carlo simulation''. Such (Monte-Carlo) samples will allow us to estimate $p^*(\bX)$ and subsequently estimate $\phi^*_\alpha(\bX)$ if desired. Clearly, this process introduces a second layer of randomness in addition to the randomness underlying the data sample, which must be taken into account in order to draw valid inference. Since these two layers play separate roles in the statement of validity guarantees in the next section, we need to keep the notation relatively heavy, as it explicitly shows that the true $p$-value of the test depends on the observed data $\bX$. When only the second layer of randomness (induced by sampling $B_i$) is of interest, we use $\bx$ to indicate that our observations are fixed and statements are conditional on the fixed observations $\bx$.

\begin{rem}\label{rem:exch}
To find a distribution $F_0(\bX)$ such that~\eqref{eq:p_val_true} stochastically dominates the continuous uniform distribution on $[0,1]$, one typically aims to find a distribution such that for a sequence of $m$ independent and identically distributed samples $\bY_i \sim F_0(\bX)$ it holds that
\[
T(\bX),T(\bY_1),\dots,T(\bY_m) \text{ are exchangeable for all }F\in\Omega_0 \ .
\]
This exchangeability then motivates the computation of a $p$-value~\eqref{eq:p_val_true}. Note that if the null class $\Omega_0 = \{F_0\}$ is a singleton (like in the scan statistic example of Section~\ref{sec:motex}) then naturally one may choose $F_0(\bX) = F_0$. In classical permutation tests, $F_0(\bX)$ corresponds to a uniform sample from the permutation distribution of $\bX$. In common bootstrap or plug-in approaches \citep{Robins2000, Beran1988} the distribution $F_0(\bX)$ may correspond to a surrogate null distribution (for example with estimated parameters) for which the above exchangeability may be only asymptotically valid.
\end{rem}

\subsection{The risks of using a predetermined fixed sample size}\label{sec:fixed-sample}

To circumvent the need to address the randomness induced by Monte-Carlo sampling, one may opt to use an extraordinarily large sample from $G^*(\bx)$ to approximate $p^*(\bx)$. In that case, it may be defensible to disregard the randomness of the Monte-Carlo layer entirely, in view of the law of large numbers. However, one needs to keep in mind that the size of the sample may need to be larger than perhaps commonly used.

For example, suppose that we wish to test at a significance level of $\alpha = 0.05$. Let our observations be fixed such that $p^*(\bx) = 0.06$. Now, suppose we use 1000 samples from $G^*(\bx)$. Following prior literature (for example, \cite{Phipson2010, North2002, Lehmann2005}) we consider the upward-biased\footnote{The problems illustrated in this section do not stem from this induced bias, and for the unbiased estimator akin to~\eqref{eq:p-est-intro-simple} the illustrated problems are equally present.} estimator for the $p$-value: 
\begin{equation}\label{eq:p-est-fixed-biased}
\hat p(\bX) = \frac{1 + \sum_{i=1}^{1000} B_i}{1001} \ ,
\end{equation}
where $B_i$ is defined as in~\eqref{eq:mc_samples}. In this case, the probability that this fixed-sample approach mistakenly implies rejection of the null (through producing a $p$-value below $\alpha$) is approximately $0.078$. Contrasting this with the planned type-I error of $\alpha=0.05$, we see that (in this setting) the fixed-simple approach increases the originally planned probability of type-I errors by a factor over $1.5$. Moreover, this inflation of the type-I error is \emph{entirely} due to the randomness of the calibration process, \emph{not} due to extreme fluctuations in the original data sampling process which is held fixed here. Whether this is acceptable may depend on the context, but we believe such anti-conservativeness, solely induced by the randomness of calibration, may be unacceptable in many contexts.

The observation above may be surprising for the unsuspecting reader; inference based on the $p$-value estimator as in~\eqref{eq:p-est-fixed-biased} is typically regarded as ``preserving type-I error control'' regardless of the sample size used \citep{Phipson2010, North2002, Ewens2003, Lehmann2005, Hemerik2018}. This is correct, but only when this error is averaged out over the distribution of the null distribution of the $p$-values. Here, we instead keep the $p$-value fixed (or rather the observations $\bx$ fixed), which offers a different perspective on the notion of ``validity'' of our inference. We discuss this distinction in detail in Section~\ref{sec:validity}; essentially the produced inference and $p$-value leads to ``unconditionally'' valid inference (meaning unconditional on the data, which we formalize in Definition~\ref{def:uncond-phi} and~\ref{def:uncond-p}) but may have potentially undesirable ``conditional'' validity (Definition~\ref{def:cond-phi},~\ref{def:cond-phi-power} and~\ref{def:cond-p}). 

\subsection{The risks of estimating the test result but reporting a \texorpdfstring{$p$-value}{p-value}}\label{sec:goals}

To set the stage for sample size recommendations, we first explicitly state the information goals the analyst may have when reporting on tests calibrated by Monte-Carlo simulation. Broadly, we can split these as follows:
\begin{itemize}
\item An estimate for the accept/reject decision at a (single) fixed predetermined significance level $\alpha$, i.e. an estimate for $\phi^*_\alpha(\bX)$ in~\eqref{eq:phi}.
\item An estimate for the $p$-value $p^*(\bX)$ in~\eqref{eq:p_val_true}. This probability corresponds to the minimum significance level $\alpha$ at which the test would have been rejected. In other words, a quantity interpretable as a $p$-value.
\end{itemize}
As far as the authors are aware, current related literature focusses mainly on providing sample size recommendations when the analyst wishes to only report an estimate for the accept/reject decision $\phi^*_\alpha(\bX)$. Such recommendations typically depend on the strength of the required validity guarantees of $\phi^*_\alpha(\bX)$ by the analyst. However, for various reasons \citep{Wasserstein2016, Greenland2016, Aguinis2021} analysts may desire to report more than only this dichotomized $p$-value $\phi_\alpha^*(\bX)$. 

Most software implementations of existing methodology which provide meaningful estimates of $\phi^*_\alpha(\bX)$ also provide an estimate for the $p$-value $p^*(\bX)$. Unfortunately, the latter comes with much fewer guarantees than may be expected by the analyst, and may possibly be misinterpreted. To illustrate this, we show how some of these approaches additionally report on $p$-value estimates which may not have the stability or carry the degree of information the analyst may expect from them. 

\begin{figure*}[ht]
\centering
\includegraphics[width=\linewidth]{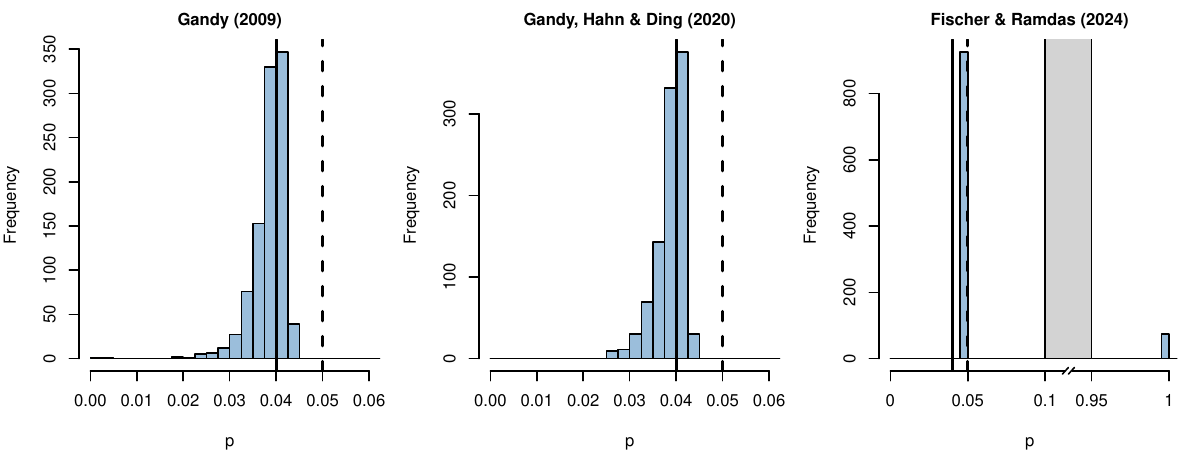}
\caption{Histograms of estimates of the $p$-value following three sequential procedures, with true $p$-value fixed at $p^*(\bx) = 0.04$. Each method was run $10^3$ times independently, and tuned as showcased by the original authors. Detailed descriptions of these methodologies are included in Section~\ref{sec:existing}. 
The method of \cite{Gandy2009} is tuned for $\alpha = 0.05$ and $\varepsilon = 10^{-3}$. 
The method of \cite{Gandy2020} is tuned for $\alpha = 0.05$, $\varepsilon = 10^{-3}$ and default significance buckets (Equation (2) in the original reference).
The method of \cite{Fischer2024} (with reference to the original work: using the ``Binomial mixture strategy'' including stopping for futility) is tuned with $c = 0.9\alpha$ and $\alpha = 0.05$. The histograms for the methods of \cite{Gandy2009} and \cite{Gandy2020} are restricted to $[0,0.06]$. The histogram of \cite{Fischer2024} is plotted with a mid-truncated $x$-axis for readability; for this method no $p$-value estimates lie in the interval $[0.1,0.95]$.} \label{fig:intro-example}
\end{figure*}

We consider a simulation setting where the observations $\bx$ are held fixed such that $p^*(\bx) = 0.04$ is the true $p$-value. We use methods that control the number of Monte-Carlo samples, and that are tuned to provide a meaningful estimate of $\phi^*_\alpha(\bx)$ at $\alpha = 0.05$. Histograms of the $p$-value estimates for three approaches are plotted in Figure~\ref{fig:intro-example}. 

Figure~\ref{fig:intro-example} highlights how the validity and usefulness of the ensuing $p$-value estimates following the three methods vary, where the methodologies are tuned as the original authors have showcased them. We stress that the \emph{perceived} shortcomings of the $p$-value estimates by these methods are no shortcomings of the methodologies themselves! These methods are designed to provide reliable estimates of $\phi^*_\alpha(\bx)$ and not of $p^*(\bx)$. With that said, there might be a temptation for the analyst to interpret the $p$-value estimates arising from these methods in an inadequate (and possibly misleading) way. Ultimately, this might steer analysts from using these methods altogether. We briefly comment on the results of the three approaches:

\begin{itemize}
\item \cite{Gandy2009} provides estimates for the $p$-value such that they result in a correct estimate for $\phi^*_{\alpha}(\bx)$ with probability $1-\varepsilon$ (over the Monte-Carlo sampling process). However, the $p$-values estimates themselves may be considerably underestimating the true $p$-value, such that an analyst over-interpreting these and reporting such estimates may falsely hint that, had the analyst chosen a much smaller value of $\alpha$ prior to the experiment, they would have rejected the null hypothesis as well.
\item \cite{Gandy2020} extends the approach of \cite{Gandy2009}. The method employs (pre-specified) overlapping significance intervals. The method identifies the significance interval in which the true $p$-value lies with probability $1-\varepsilon$. However, like the method \cite{Gandy2009} the analyst should not over-interpret the $p$-values estimates themselves; these may still be considerably underestimating the true $p$-value.
\item \cite{Fischer2024} produces $p$-values, but these carry no additional information other than the binary decision value $\phi^*_{\alpha}(\bx)$. Conversely to \cite{Gandy2009} or \cite{Gandy2020}, there is no simple tuning parameter that relates to the probability of the $p$-value estimate resulting in a wrong estimate of $\phi^*_\alpha(\bx)$.
\end{itemize}
In Section~\ref{sec:existing} we survey these and other approaches in more detail.

\subsection{Outline}
The rest of the paper is structured as follows. 
In Section~\ref{sec:validity} we discuss what validity guarantees may be desirable. These are dependent on the goals of the analyst. We also show how validity guarantees previously presented in the literature connect to each other. Having set the stage in terms of outcome goals and validity guarantees, we survey existing approaches in Section~\ref{sec:existing}.
We propose a novel methodology for sequential $p$-value estimation in Section~\ref{sec:novel}. Finally in Section~\ref{sec:recommend} we use our survey to give recommendations for best practices to the analyst; both in terms of selection of methodology, as well as in the manner of reporting.

\section{Validity of Inference}\label{sec:validity}

The discussion surrounding the results in Figure~\ref{fig:intro-example} highlights how current approaches in the literature aim to provide different validity guarantees, and a valid estimate of $\phi^*_\alpha(\bx)$ does not imply the $p$-value estimate it is based on it carries any additional information. We start by discussing sensible validity guarantees of the test outcome estimate $\phi^*_\alpha(\bx)$ and then move on to define suitable validity guarantees for $p$-value estimates.

\subsection{Validity of the test outcome estimate \texorpdfstring{$\hat\phi_\alpha(\bx)$}{hat phi(bx)}}\label{sec:validity-phi}
Consider an estimator $\hat\phi_\alpha(\bX)$ of the testing outcome $\phi^*_\alpha(\bX)$. Recall that $\hat\phi_\alpha(\bX)$ is random through both the data $\bX$ as well as the Monte-Carlo sampling. For some fixed prescribed choice $\alpha$, two distinct notions of validity are commonly discussed in the literature. For ease of reference we state these here as definitions.

\begin{defi}[Unconditionally valid $\hat\phi_\alpha(\bX)$]\label{def:uncond-phi}
An estimator $\hat \phi_\alpha(\bX)$ for the testing decision $\phi^*_\alpha(\bX)$ is \emph{unconditionally valid} if, uniformly over all null hypothesis models $F\in\Omega_0$, it has type-I error bounded by the target level of significance, i.e.
\[
\forall_{F\in\Omega_0} : \mathbb{P}_{\bX \sim F}\left( \hat \phi_{\alpha}(\bX) = 1 \right) \leq \alpha \ .
\]
\end{defi}

The above validity statement can be seen as a baseline requirement for inference based on Monte-Carlo simulation, but solely guaranteeing this property may not suffice for the analyst. In an extreme case, an estimator $\hat\phi_\alpha(\bX)$ can easily satisfy the condition above and may be completely independent of the data $\bX$, corresponding to a powerless test.

We refer to the property in the above definition as ``unconditional'', as the probability above is not conditional on the observed data $\bX$, but rather takes into account both relevant layers of randomness: the randomness of the data as well as that of the Monte-Carlo samples. This ensures that the overall type-I error of the test calibrated by Monte-Carlo simulation is at most our prescribed significance. Conversely, a guarantee conditional on the true test outcome $\phi^*_\alpha(\bx)$ can be stated through the resampling risk (see, for example, \cite{Jockel1984, Gandy2009, Davidson2000}):

\begin{defi}[$\varepsilon$-bounded resampling risk of $\hat\phi_\alpha(\bX)$]\label{def:cond-phi}
For an estimator $\hat \phi_\alpha(\bX)$ define the resampling risk as
\[
\text{RR}_\alpha\Big(\hat\phi_{\alpha} ; \bx\Big) \defeq \mathbb{P}\left( \hat\phi_\alpha(\bX) \neq \phi^*_\alpha(\bX) \hspace{0.1cm}\Big\vert\hspace{0.1cm}  \bX=\bx \right)\ .
\]
We say estimator $\hat\phi_{\alpha}$ has \emph{$\varepsilon$-bounded resampling risk} at level $\alpha$ if
\[
\sup_{\bx\in\mathcal{X}} \text{RR}_\alpha\Big(\hat\phi_{\alpha} ; \bx\Big) \leq \varepsilon\ .
\]
\end{defi}

In the above definition only the randomness from the Monte-Carlo sampling plays a role. Some related literature positions the resampling risk as a property of a $p$-value estimator $\hat p(\bx)$. Although this is not incorrect, we refrain from such statements since it may be interpreted too broadly: a sharp bound on the resampling risk guarantees nothing for the $p$-value estimator apart from the sign of $(\hat p(\bx)-\alpha)$ --- which is precisely the information that $\phi_\alpha^*(\bx)$ provides. To avoid any confusion we state this validity guarantee as pertaining the estimated test outcome $\hat \phi_{\alpha }(\bx)$. This validity property is ``conditional'' on the data through its dependency on $\bx$. 

While both of these two validity guarantees are sensible and may be desirable in practice, existing literature typically focuses on one notion without discussing the other. Here we connect these two distinct notions of validity.

A first question one may be interested in is if one of the validity notions implies the other. Since unconditional validity can be satisfied by a powerless test while it cannot satisfy $\varepsilon$-bounded resampling risk, unconditional validity cannot imply conditional validity. It is less clear if the converse holds, but one can show it does not hold in generality. Instead, conditional validity implies unconditional validity \emph{only} under some additional assumptions on $\hat\phi_\alpha(\bx)$. For brevity we defer a detailed discussion on what types of assumptions would be needed for such an implication in Section~\ref{sec:connecting-validities-phi}. 

Nevertheless, of practical importance is the observation that any estimator with $\varepsilon$-bounded resampling risk is guaranteed to be ``nearly'' unconditional valid:
\begin{lem}\label{lem:rr-implies-near-uncond}
Let $\hat\phi_\alpha(\bX)$ be an estimator for the testing outcome $\phi^*(\bX)$ with $\varepsilon$-bounded resampling risk as in Definition~\ref{def:cond-phi}. Then
\[
\forall_{F\in\Omega_0} : \mathbb{P}_{\bX \sim F}\left( \hat \phi_{\alpha}(\bX) = 1 \right) \leq \alpha + \varepsilon \ .
\]
\end{lem}
In other words, an estimator with $\varepsilon$-bounded resampling risk has an overall type-I error rate inflated by \emph{at most} $\varepsilon$. For the analyst unsure which validity notion is important, the above observation may be helpful in their decision. We stress that the result above provides an \emph{upper} bound for the overall type-I error, and under additional assumptions (going beyond the $\varepsilon$-bounded resampling risk of Definition~\ref{def:cond-phi}) on the estimator $\hat\phi_\alpha(\bX)$ one may show it satisfies the sharper bound in Definition~\ref{def:uncond-phi}.

To further highlight the connections between the two notions of validity, for the same reasons underlying Lemma~\ref{lem:rr-implies-near-uncond}, one can \emph{transform} an estimator with $\varepsilon$-bounded resampling risk to an unconditionally valid estimator. However, there is a caveat to this result.
\begin{lem}\label{lem:phi-cond-to-uncond}
Let $\alpha\in(0,1)$ be the desired significance level. Let $\varepsilon \in (0,\alpha)$ and let $\hat\phi_{\alpha-\varepsilon}(\bX)$ be an estimator for the testing outcome at a stringer significance level $\alpha-\varepsilon$ such that at that level the estimator has $\varepsilon$-bounded resampling risk (Definition~\ref{def:cond-phi}). That is, suppose that $\sup_{x\in\cX}\text{RR}_{\alpha-\varepsilon}(\hat\phi_{\alpha-\varepsilon} ; \bx) \leq \varepsilon$.
Construct an estimator for $\phi^*_\alpha(\bx)$ as
\[
\tilde \phi_\alpha(\bx) = \hat\phi_{\alpha-\varepsilon}(\bx) \ .
\]
Then $\tilde\phi_\alpha(\bX)$ has unconditionally valid type-I error at the significance level of interest (Definition~\ref{def:uncond-phi}), i.e.
\[
\forall_{F\in\Omega_0} : \mathbb{P}_{\bX \sim F}\left( \tilde\phi_{\alpha}(\bX) = 1 \right) \leq \alpha\ .
\]
\end{lem}
The proof is a direct consequence of Lemma~\ref{lem:rr-implies-near-uncond} by substitution of $\alpha = \alpha -\varepsilon$.

Although the newly constructed $\tilde\phi_\alpha(\bX)$ of Lemma~\ref{lem:phi-cond-to-uncond} also retains a resampling risk control, there is a caveat. One must carefully note this bound is with respect to $\phi^*_{\alpha-\varepsilon}(\bX)$, i.e.
\[
\forall_{\bx\in\cX} : \P{\tilde \phi_\alpha(\bX) \neq \phi^*_{\alpha-\varepsilon}(\bX) \given \bX = \bx} \leq \varepsilon \ .
\]
The mismatched significance levels of $\tilde\phi_\alpha(\bX)$ and $\phi_{\alpha-\varepsilon}^*(\bX)$ in the above bound are undesirable, such that the practical value of the above bound for the construction $\tilde\phi_\alpha(\bx)$ may be harder to appreciate.

The above lemma gives perhaps some hope that a construction could exists, based on an estimator with bounded resampling risk, such that the newly constructed estimator has \emph{both} unconditional validity as well as $\varepsilon$-bounded resampling risk. We illustrate in Section~\ref{sec:connecting-validities-phi} why this is not possible for simple constructions, and assumptions going beyond $\varepsilon$-bounded resampling risk are necessarily to proceed further.

Desiring a prescribed bound on the resampling risk may come at potentially large computational costs, since bounding the resampling risk becomes increasingly harder if the unknown true $p$-value $p^*(\bx)$ is close to $\alpha$. Furthermore, it is impossible to tightly bound the resampling risk in finite time when $p^*(\bx) = \alpha$, and in that specific scenario, methods designed to ensure tight bounds never stop sampling as a consequence. On these grounds, one may opt to refrain from controlling the resampling risk, but rather consider a weaker conditional criterion (see our discussion on \cite{Silva2018} in Section~\ref{sec:bc}). One such criterion previously considered is the (conditional) power loss of the test:

\begin{defi}[$\varepsilon$-bounded power loss of $\hat\phi_\alpha(\bX)$]\label{def:cond-phi-power}
An estimator $\hat \phi_\alpha(\bX)$ for the testing decision $\phi^*_\alpha(\bX)$ has \emph{$\varepsilon$-bounded power loss} if for all $\bx$ such that $\phi^*_\alpha(\bx) = 1$ the probability of failing to reject the null hypothesis is bounded by $\varepsilon$. Formally,
\[
\sup_{\bx\in\mathcal{X}_\alpha} \P{ \hat\phi_\alpha(\bX) = 0 \given \bX=\bx} \leq \varepsilon\ ,
\]
where $\mathcal{X}_{\alpha} \defeq \left\{\bx\in\mathcal{X}: \phi^*_\alpha(\bx) = 1 \right\}$.
\end{defi}

Clearly, a bound on the resampling risk bounds the power loss as well. Since the above guarantee can be trivially satisfied by using $\hat\phi_\alpha(\bx) = 0$ it is clear that additional validity requirements are necessary for estimators to be meaningful.

While the validity properties discussed ensure attractive properties of the inference following the outcome estimators, one should note that the test $\hat\phi_\alpha(\bX)$ is still randomized. This means that guarantees apply for a \emph{single usage} of the testing procedure, and in order to retain such guarantees the analyst is not allowed to repeat the testing procedure multiple times on the same data.

\subsection{Validity of the \texorpdfstring{$p$-value estimate $\hat p(\bx)$}{p-value estimate hat p(x)}}\label{sec:p_validity}

We now discuss sensible validity guarantees for a $p$-value estimate, akin to the guarantees for an estimate of $\phi^*_\alpha(\bx)$ as discussed in the previous section. A first natural requirement for a $p$-value estimate is (unconditionally on the observed data) stochastic domination of the uniform distribution under the null hypothesis, essentially generalizing Definition~\ref{def:uncond-phi} to hold for all values of~$\alpha$:

\begin{defi}[Unconditionally valid $\hat p(\bX)$]\label{def:uncond-p}
An estimator $\hat p(\bX)$ for the $p$-value $p^*(\bX)$ is \emph{unconditionally valid} if, uniformly over all models $F\in\Omega_0$ it stochastically dominates the uniform distribution:
\[
\forall_{F\in\Omega_0} \forall_{t\in[0,1]} : \mathbb{P}_{\bX \sim F}\left( \hat p(\bX) \leq t\right) \leq t \ .
\]
\end{defi}

This requirement ensures that the overall type-I error of a test based on the $p$-value estimate produced through Monte-Carlo sampling is at most our prescribed significance, taking into account both the randomness underlying our data sample as well as the randomness due to the Monte-Carlo calibration. This requirement can be seen as a baseline property of an estimator of the $p$-value. Estimators satisfying the above can be be easily constructed even when ignoring the data (e.g., take $\hat p(\bX)=1$). Therefore, this guarantee is, in a sense, a minimal requirement for any $p$-value estimator.

Next, we construct a validity requirement for an estimator $\hat p(\bX)$ with a guarantee conditional on the observed data, i.e. akin to the resampling risk guarantee for the estimator of $\phi^*_\alpha(\bX)$ in Definition~\ref{def:cond-phi}. Recall that the $p$-value corresponds to the smallest significance value for which the analyst would have rejected the null. Ideally, our estimate would match this precisely, but such a requirement is only possible to satisfy with infinite Monte-Carlo samples as it requires exact knowledge of $p^*(\bx)$. Instead we focus only (as a baseline requirement) on bounding type-I errors induced by the Monte-Carlo sampling --- as is traditional in statistical inductive reasoning. This is similar in spirit to the bound on power loss of Definition~\ref{def:cond-phi-power}, except instead of bounding type-II errors we instead bound the type-I errors induced by the Monte-Carlo sampling --- and rather than fixing a significance level $\alpha$, we must do so over all potential significance levels.

For a $p$-value estimate $\hat p(\bx)$, a type-I error induced by the Monte-Carlo sampling process could arise when a significance value exists for which the estimate $\hat p(\bx)$ implies the null can be rejected, while the exact $p$-value $p^*(\bx)$ does not. A sensible validity requirement would then impose an upper bound on the probability that such an undesirable event happens by some (small) prescribed probability $\varepsilon > 0$. Note that this implies that we need to produce a $p$-value estimate which is larger than the true $p$-value with high probability; i.e. an upper confidence limit. However, estimators which constitute such an upper confidence limit come with the interpretation that, with high probability, such estimates do not overestimate the significance of the observed statistic $T(\bx)$. Because of this, such upper confidence limits are a useful concept within our context, and it is effective to descriptively label estimators which satisfy this property. We introduce this concept here as the risk of overestimated significance (ROS):

\begin{defi}[$\varepsilon$-bounded risk of overestimated significance (ROS)]\label{def:cond-p}
For an estimator $\hat p(\bX)$ of $p^*(\bX)$ define the risk of overestimated significance (ROS) as
\[
\text{ROS}\Big(\hat p, \bx\Big) \defeq \mathbb{P} \Big( \hat p(\bX) < p^*(\bX) \Big\vert\hspace{0.1cm} \bX=\bx \Big)\ .\]
We say estimator $\hat p$ has \emph{$\varepsilon$-bounded risk of overestimated significance (ROS)} if
\[
\sup_{\bx\in\mathcal{X}} \text{ROS}\Big(\hat p, \bx\Big) \leq \varepsilon\ .
\]
\end{defi}
Note that in the probability statement above, the randomness is due to the Monte-Carlo sampling only. The powerless estimator $\hat p(\bX)=1$ will also satisfy the above definition. This is where the above conditional property differs from the $\varepsilon$-bounded resampling risk of Definition~\ref{def:cond-phi}, which cannot be satisfied for trivial estimators of $\phi^*_\alpha(\bx)$.

The two validity guarantees above play distinct roles (akin to the distinct nature of Definition~\ref{def:uncond-phi} and Definition~\ref{def:cond-phi} in the previous section). The unconditional validity in Definition~\ref{def:uncond-p} guarantees the complete testing process does not have an inflated type-I error. Additionally, the conditional validity guarantee in Definition~\ref{def:cond-p} ensures that the type-I errors due \emph{solely} to the randomness of the Monte-Carlo simulation-layer can be bounded apart from the overall error, at a potentially much smaller order than the overall type-I error budget $\alpha$; this may be desirable in many practical settings.

Importantly, both validity guarantees Definition~\ref{def:uncond-p} and Definition~\ref{def:cond-p} are independent of the significance level $\alpha$ at which the analyst draws their inference. This allows the $p$-value estimate to admit its usual interpretation --- as the smallest significance level at which we would have rejected the null hypothesis --- which is also independent from the significance level chosen by the analyst.

If $\hat p(\bX)$ has $\varepsilon$-bounded ROS then we can use such estimators to easily construct a new estimator $\tilde p(\bX)$ that satisfies \emph{both} Definition~\ref{def:uncond-p} and Definition~\ref{def:cond-p}:

\begin{lem}\label{lem:p-cond-to-uncond}
Let $\hat p(\bX)$ be an estimator for $p^*(\bX)$ for which $\sup_{\bx\in\cX}\text{ROS}\big(\hat p, \bx\big) \leq \varepsilon$. Then we can construct an estimator 
\[
\tilde p(\bX) \defeq \hat p(\bX) + \varepsilon \ ,
\]
such that both $\sup_{\bx\in\mathcal{X}}\text{ROS}\big(\tilde p, \bx\big) \leq \varepsilon$ as well as for all $t\in[0,1]$ and for all $F\in\Omega_0$:
\[
\forall_{F\in\Omega_0} \forall_{t\in[0,1]}: \mathbb{P}_{\bX \sim F}\left( \tilde p(\bX) \leq t \right) \leq t \ .
\]
\end{lem}
The proof is deferred to Section~\ref{sec:p-cond-to-uncond}. Again, and similar to our final remarks in Section~\ref{sec:validity-phi}, since the estimator $\tilde p(\bX)$ from the above lemma is still random, the above validity guarantees only apply if the analyst uses the testing procedure only once. While the guarantee on the ROS controls the type-I errors that may arise from instabilities in the estimate $\hat p(\bx)$, ensuring some reproduction-stability in the range of significance levels at which the analyst may reject the null hypothesis, the analyst should be mindful that the reported $p$-value estimate still relies on randomization that can only be exactly reproduced when randomization seeds are shared.

\section{Existing Approaches}\label{sec:existing}
This section surveys existing approaches which draw inference from a test calibrated by Monte-Carlo sampling. We reflect these approaches back to the inference goals of Section~\ref{sec:goals} and their respective validity guarantees of Section~\ref{sec:validity}. To avoid computational inefficiencies, we have a particular focus on approaches based on sequential sampling methods, deciding on-the-fly on how many Monte-Carlo samples are needed to draw inference. We group approaches based on our perception of their underlying design principles. 

\begin{figure*}[ht]
\centering
\begin{subfigure}{0.45\linewidth}
\includegraphics[width=\linewidth]{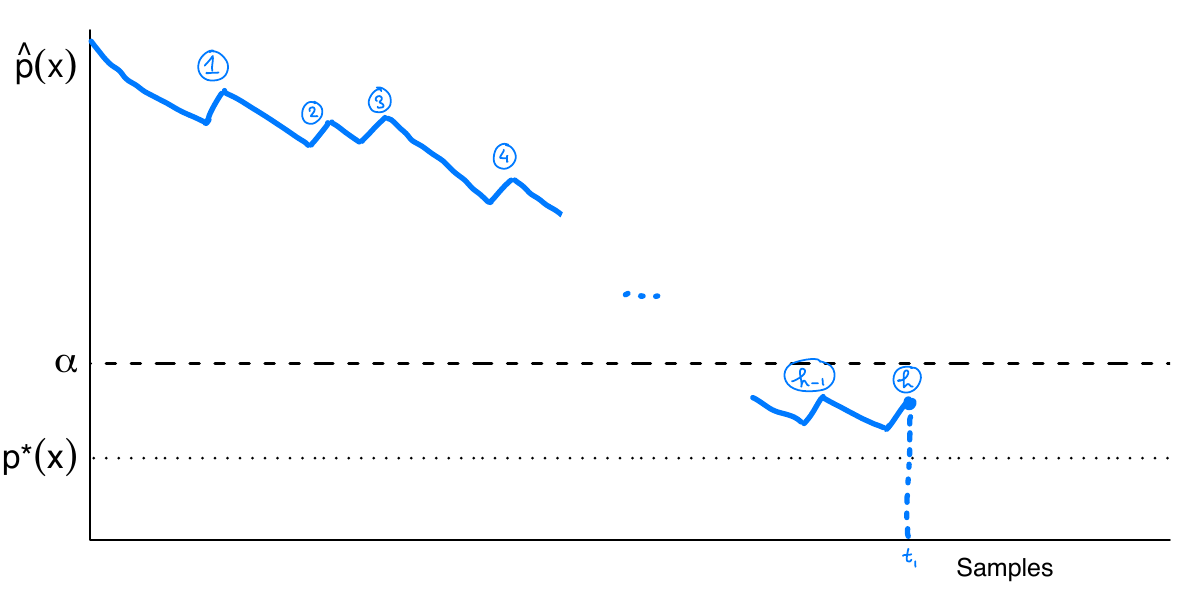}
\caption{\cite{Besag1991}.}\label{fig:sketch-besag}
\end{subfigure}
\begin{subfigure}{0.45\linewidth}
\includegraphics[width=\linewidth]{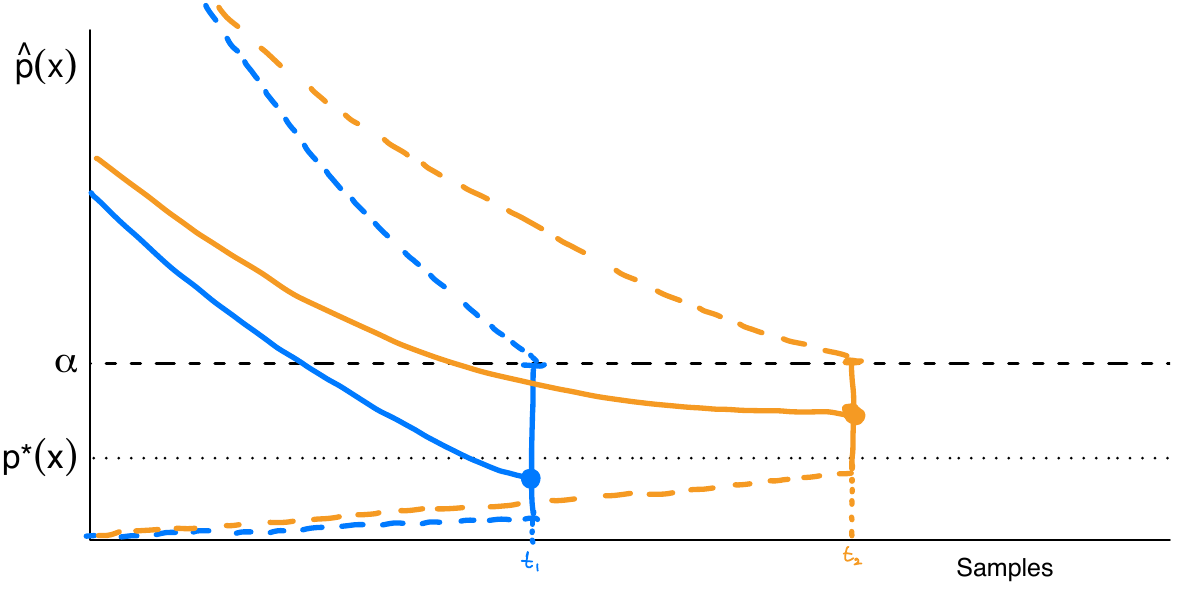}
\caption{\cite{Gandy2009}.}\label{fig:sketch-gandy}
\end{subfigure}
\begin{subfigure}{0.45\linewidth}
\includegraphics[width=\linewidth]{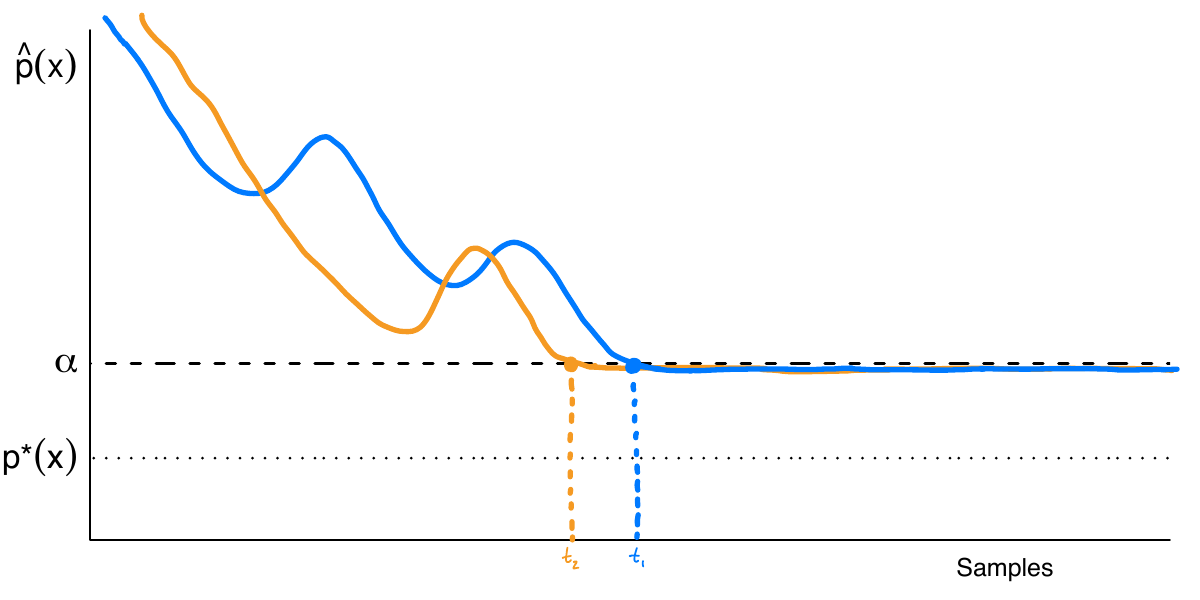}
\caption{\cite{Fischer2024}.}\label{fig:sketch-fischer}
\end{subfigure}
\begin{subfigure}{0.45\linewidth}
\includegraphics[width=\linewidth]{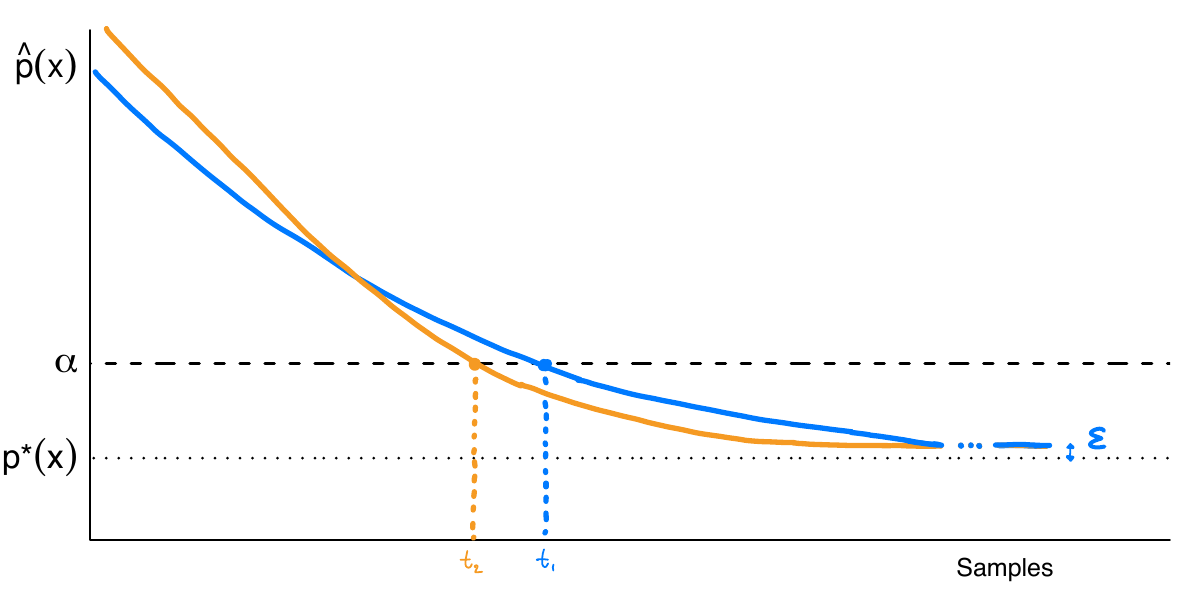}
\caption{Our own proposal (Definition~\ref{def:proposal}).}\label{fig:sketch-own}
\end{subfigure}
\caption{Sketches showing typical behaviour of the running $p$-value estimate of a selection of the surveyed methodologies in Section~\ref{sec:existing}. Note that not all methodologies shown should be monitored as such. The sketches serve to illustrate the methodologies and showing connections only, and are further explained at their place of reference.}\label{fig:cartoons}
\end{figure*}

For a selection of the methodologies, a sketch showcasing how the $p$-value estimates typically evolve over time and lead to a final estimate is given in Figure~\ref{fig:cartoons}. The goal of this figure is to provide a qualitative illustration of the inner working, outcomes, and show connections between the methods, rather than providing quantitative value.

\subsection{Fixed predetermined samples}\label{sec:fixed-predet}
The focus of our work is on approaches which prudently spend computational resources, adapted to the goals and validity requirements of the analyst. As already illustrated in Section~\ref{sec:fixed-sample} classical predetermined sampling strategies are typically unsuitable for such goals. Nevertheless, for completeness we briefly contrast here approaches using a predetermined sample size. A seemingly sensible $p$-value estimator using (predetermined) $m$ i.i.d. samples $B_1,B_2,\dots, B_m$ from $G^*(\bx)$ as in~\eqref{eq:mc_samples}, is given by its maximum likelihood estimator:
\begin{equation}\label{eq:p_estimator_naive}
\hat p_{m,\text{Naive}}(\bX) \defeq \frac{\sum_{i=1}^m B_i}{m} \ .
\end{equation}
As noted by other authors, this straightforward estimator, while ubiquitous in practice, does not even satisfy the baseline unconditional requirement of Definition~\ref{def:uncond-p} \citep{Phipson2010, North2002, Lehmann2005}. Instead a biased (but asymptotically consistent) estimator for the $p$-value using a fixed predetermined sample is given by
\begin{equation}\label{eq:p_estimator_bias}
\hat p_{m,\text{Biased}}(\bX) \defeq \frac{1+\sum_{i=1}^m B_i}{m + 1} \ .
\end{equation}
For large $m$ the difference is minor, but the slight bias endows the estimator with unconditional validity as in Definition~\ref{def:uncond-p} regardless of the choice of $m$. Contrarily, conditional validity properties depend highly on both the choice of $m$, as well as the unknown value of $p^*(\bx)$, and choosing $m$ without knowledge of $p^*(\bx)$ requires taking a pessimistic stance, incurring large computational costs.

For small $m$, the added bias and discreteness of the estimator~\eqref{eq:p_estimator_bias} may substantially affect testing power. As proposed in \cite{Racine2007}, one can refine the estimator through external randomization as:
\[
\hat p_{m,\text{RM}}(\bX) \defeq \frac{U+\sum_{i=1}^m B_i}{m + 1} \ ,
\]
where $U$ is a single uniform random variable supported on $[0,1]$, independent from the data and the random samples $B_i$. This makes the estimated $p$-value exactly uniform for any choice of $m$ (i.e. Definition~\ref{def:uncond-p} holds exactly) at the cost of additional sources of randomness. The latter further complicates securing conditional validity guarantees.

\subsection{Estimating sample sizes based on preliminary samples}\label{sec:prelim-comp}
In \cite{Andrews2000, Andrews2001} a three-step approach to determine a suitable sample size for the $p$-value estimator~\eqref{eq:p_estimator_naive} is proposed. It is most meaningful when the asymptotic distribution $F_\infty$ of the test statistic is $T(\bX)$ is available. The authors take a different stance with respect to desired behaviour of the $p$-value estimator than we have discussed so far. Instead of (explicitly) aiming for guarantees relating to validity of ensuing inference, the authors aim to select the required number of samples $m$ which guarantees that (conditional on the observations $\bx$) the relative distance between the $p$-value estimator~\eqref{eq:p_estimator_naive} and the true $p$-value is bounded with high probability:
\[
\P{\frac{\abs{\hat p_{m,\text{Naive}}(\bX) - p^*(\bX)}}{p^*(\bX)} \leq d \given \bX = \bx} \approx 1-\tau \ ,
\]
for pre-specified $d$ and $\tau$. The above condition is not guaranteed exactly by the procedure, but is valid asymptotically as $d\to 0$, based on the asymptotic normality result conditional on $\bx$ and as $m \to \infty$:
\begin{equation}\label{eq:asymp-naive}
\sqrt{m}\Big(\hat p_{m,\text{Naive}}(\bx) - p^*(\bx)\Big) \to \cN\Big(0,p^*(\bx)(1-p^*(\bx))\Big) \ ,
\end{equation}
where the difficulty in leveraging the above result obviously lies in the absence of knowledge of $p^*(\bx)$. The proposed procedure consists of three steps, aiming to approximate the above variance to inform of a reasonable sample size. First, a crude estimate of the variance in~\eqref{eq:asymp-naive} is computed\footnote{If $F_\infty$ is unavailable then the pessimistic maximum value $\omega_0 = 1/4$ can be used, which renders subsequent steps irrelevant and will pessimistically recommend a sample size based solely on $d$ and $\tau$. } as $\omega_0 = F_\infty(T(\bx))/(1-F_\infty(T(\bx)))$ and a preliminary sample of size $m_1 = z_{1-\tau/2}^2\omega_0/d^2$ is sampled from $G^*(\bx)$. Second, this sample is used to more accurately estimate the variance in~\eqref{eq:asymp-naive} by $\hat\omega = \hat p_{m_1,\text{Naive}}(\bx)(1-\hat p_{m_1,\text{Naive}}(\bx))$, and a refinement for the required sample size is computed as $m_2 = z_{1-\tau/2}^2\hat\omega/d^2$. Third, the final sample size is determined as $\max(m_1,m_2)$, where additional samples from $G^*(\bx)$ are obtained if necessary, and the final estimate computed based on~\eqref{eq:p_estimator_naive}.

\subsection{Sequential sampling and stopping early if \texorpdfstring{$p^*(\bx)$}{p*(x)} is large}\label{sec:bc}
To be best of our knowledge, the oldest methodology to sequentially sample in this context is introduced in \cite{Besag1991}. This is a heuristic approach which collects Monte-Carlo samples sequentially, aiming to keep sampling and provide accurate estimates of $p^*(\bx)$ if it is small while stopping early if $p^*(\bx)$ is large. The methodology is based on two tuning parameters $h$ and $M$ which work as follows: sequentially, the Monte-Carlo samples $B_i$ are computed until either $h$ occurrences of $B_i = 1$ are observed, or until $M$ samples are drawn. The former stopping criterion (heuristically) encodes the idea that one should stop sampling if the underlying $p$-value is suspected to be large. If $p^*(\bx)$ is small, then the procedure is likely to keep sampling until its maximum $M$, whereas for large $p^*(\bx)$ values the threshold $h$ is reached earlier. Letting $\ell$ denote the index at which the procedure is stopped, the $p$-value estimate of this methodology is given by
\[
\hat p_\text{BC}(\bX) \defeq \begin{cases} \frac{h}{\ell} & \text{if } \ell < M ; \\ \frac{1+\sum_{i=1}^M B_i}{M+1} &\text{otherwise} \ .\end{cases} 
\]
The ensuing $p$-value is unconditionally valid (i.e. as in Definition~\ref{def:uncond-p}), but the methodology comes with no conditional validity guarantees. See Figure~\ref{fig:sketch-besag} for a sketch of this methodology, where for illustration purposes the running $p$-value estimate is sketched, which terminates after $h$ ``jumps'' upward. Note only the estimate at termination should be used. 

In \cite{Silva2009} it is shown that for any $m$ used to construct the fixed-sample estimator~\eqref{eq:p_estimator_bias}, there exists a set of tuning parameters $h$ and $M$ such that the method by \cite{Besag1991} has the same power as inference based on~\eqref{eq:p_estimator_bias}, but based on potentially much fewer Monte-Carlo samples.

We perceive the work of \cite{Silva2018} as an extension of the methodology by \cite{Besag1991} by introducing two more tuning parameters: a parameter $t_1$ and a parameter $C_e$. The first tuning parameter limits the sampling index at which early stopping can occur; their methodology likewise stops when $h$ samples of $B_i = 1$ are observed, but only if this is done before sample $t_1$. The second tuning parameter is used as a rejection criterion; the null hypothesis is rejected if it has not stopped early, and if at the final sample $\sum_{i=1}^M B_i < C_e$. Their estimate for the testing outcome $\phi^*_\alpha(\bx)$ is therefore:
\[
\hat \phi_{\alpha,\text{SA}}(\bX) \defeq \begin{cases}0 & \text{if } \exists_{t \leq t_1} : \sum_{i=1}^t B_i \geq h \text{ or } \Big(\sum_{i=1}^{t_1} B_i < h \ , \sum_{i=1}^M \geq C_e \Big); \\ 1 &\text{Otherwise}. \end{cases}
\]
The authors show that sets of tuning parameters exist for which the ensuing inference has unconditional validity as in Definition~\ref{def:uncond-phi} as well as conditionally power bounded at prescribed level $\varepsilon$. To find these tuning parameters, two of the four parameters may be fixed and the other two found through a grid search. Note that the methodology does not come with conditional control of type-I errors, which may be substantial depending on the tuning parameters chosen. The authors also include a method to compute $p$-values based on the proposed sequential procedure, which comes endowed with unconditional validity as in Definition~\ref{def:uncond-p}, but no other guarantees are given for this estimate. A refinement of their methodology is presented, that may allow the methodology to stop earlier, particularly under the null. This refinement can stop before a collection of times $t_j$ if the running sums exceed the corresponding threshold $\sum_{i=1}^{t} B_i \geq h_{t_j}$, or drop below a corresponding threshold $\sum_{i=1}^{t} B_i \leq v_{t_j}$. The authors show how this relates to tracking a transformation of the running sum $\sum_{i=1}^t B_i$ against flat thresholds which allows finding suitable tuning parameters. These transformations require computations of cumulative densities of hypergeometric distributions, inducing a potentially large computational overhead.

\subsection{Sequential Probability Likelihood Ratio approaches}
The problem of sampling Monte-Carlo samples until a decision can be reached may naturally be addressed through a sequential probability ratio approach \citep{Wald1945}. In this context, this was first proposed pragmatically by \cite{Lock1991}, and studied in in more detail in \cite{Fay2007}. Casting the problem through this lens requires us to define a well-separated null and alternative hypothesis for the value of $p(\bx)$, i.e. for some $\delta > 0$:
\[
H_0: p^*(\bx) \leq \alpha - \delta \text{ vs. } H_1: p^*(\bx) \geq \alpha + \delta \ .
\] 
In \cite{Fay2007} a procedure is constructed such that ensuing $p$-value estimates are unconditionally valid (i.e. Definition~\ref{def:uncond-p}) leading to estimates of $\phi^*_\alpha(\bx)$ which are unconditionally valid (i.e. Definition~\ref{def:uncond-phi}). However, by the need for separation of the null and alternative a natural consequence is that conditional validity of inference (Definition~\ref{def:cond-phi}) if $p^*(\bx) \in (\alpha - \delta, \alpha + \delta)$ cannot be ensured. Without assuming $p^*(\bx)$ to be outside of this problematic region, the ensuing inference cannot be supplied with conditional validity guarantees.

\subsection{Sequential resampling-risk spending approaches} 
In \cite{Gandy2009} the authors propose a methodology to estimate $\hat\phi_\alpha(\bx)$, which sequentially samples from $G^*(\bx)$ and tests after each sample if it can be guaranteed that its corresponding estimate for the testing outcome $\hat \phi_\alpha(\bx)$ has $\varepsilon$-bounded resampling risk (as Definition~\ref{def:cond-phi}). The (prescribed) maximum resampling risk $\varepsilon$ is ``spent'' sequentially over these tests. The methodology as originally proposed is not unconditionally valid (as in Definition~\ref{def:uncond-phi}). There is also an unfortunate caveat for this methodology induced by its bounding of the resampling risk. If $\alpha = p(\bx)$ it is highly probable that the methodology never stops. Furthermore, any truncation of the method invalidates the uniform bound over the resampling risk, such that the analyst aiming to have this guarantee should not stop the methodology prematurely. However, if $p(\bx)$ is far from $\alpha$, the methodology can stop early. See Figure~\ref{fig:sketch-gandy} for a sketch of this methodology. The running $p$-value estimate is sketched with a solid line, and the confidence bands around this estimate by a dotted line (which sequentially exclude potential values of the $p$-value by spending the resampling risk $\varepsilon$). Only at termination should an estimate of $\phi_\alpha(\bx)$ be obtained. Two distinct runs of the methodology are illustrated in blue and orange.

An earlier, heuristic approach, stemming from the same ideas as those of \cite{Gandy2009} was proposed in \cite{Davidson2000}. The methodology depends on two tuning parameters $\varepsilon$ and $B_\text{min}$. It sequentially samples sets of size $B_\text{min}$ of Monte-Carlo samples. After each sampled set, it first estimates the value of $p^*(\bx)$ using $\hat p_{m,\text{Naive}}(\bx)$ from~\eqref{eq:p_estimator_naive} based on entire collection of samples collected so far. It then compares this estimate with the (predetermined) value of $\alpha$. If we observe for our estimate $\hat p_{m,\text{Naive}}(\bx) < \alpha$, then we evaluate if the null hypothesis that $p^*(\bx) \geq \alpha$ can be rejected at predetermined level $\varepsilon$. If instead we observe $\hat p_{m,\text{Naive}}(\bx) > \alpha$, then the null hypothesis $p^*(\bx) < \alpha$ is tested at level $\varepsilon$. If the conducted hypothesis test rejects its respective null, the methodology reports the final estimator $\hat p_{m,\text{Naive}}(\bx)$ as the estimated $p$-value. This heuristic methodology does not come with any validity guarantees. Contrarily to \cite{Gandy2009}, an estimator for $\phi^*_\alpha(\bx)$ based on this heuristic does not ``spend'' the resampling risk $\varepsilon$ over time due to multiplicity in testing after each sampled set. This implies the resampling risk could be substantial, particularly if $B_\text{min}$ is chosen small.

A simplified approach to that of \cite{Gandy2009}, where stopping boundaries are computed slightly conservatively but computationally much faster, is proposed in \cite{Ding2020}.

In \cite{Gandy2020} the authors propose to apply the methodology of \cite{Gandy2009} simultaneously for a pre-specified set of significance levels, i.e. estimating $\phi^*_{\alpha_i}(\bx)$ for prespecified $\{\alpha_i\}_{i=1}^k$ constructing so-called ``significance buckets'' (which must overlap), and using the Monte-Carlo samples to infer the location of the $p$-value among those significance buckets. The advantage is that the methodology is now guaranteed to eventually stop. The methodology bounds the probability of selecting the incorrect ``significance bucket'' by a prescribed probability $\varepsilon$. This implies that the upper end of the selected significance bucket provides a (granular) estimate of the $p$-value with $\varepsilon$-bounded ROS (Definition~\ref{def:cond-p}), where this granularity depends on the pre-specified significance levels. 

\subsection{Anytime-valid approaches}
In \cite{Fischer2024} an anytime-valid approach is proposed; this is a sequential approach which, unlike the previous sequential approaches described, can be stopped (or continued) at \emph{any} time while retaining an unconditional validity guarantee on the $p$-value estimate (Definition~\ref{def:uncond-p}). This permits the analyst to stop depending on the estimates of the $p$-value seen so far. Their most intricate method (referred to as the ``Binomial mixture strategy'') is an approach based around a single tuning parameter $c < \alpha$. The $p$-value estimate converges to either $\alpha$ if $p^*(\bx) < c$, and converges to 1 otherwise --- meaning that the methodology asymptotically provides $p$-values estimates which carry no more information than an estimate of $\phi^*_\alpha(\bx)$. See Figure~\ref{fig:sketch-fischer} for a sketch of this methodology. The running $p$-value estimates are sketched for illustration purposes. These can be used at any time to draw inference. Two distinct runs are sketched in blue and orange. 

The authors have considered, in follow-up work, a modification such that the methodology is suitable in a multiple-testing setting \citep{Fischer2024a}. A construction is proposed where the sequential $p$-value estimate does not converge to either $\alpha$ or 1, but rather maintains a closer adherence to the true $p$-value. These come only with an anytime-valid unconditional guarantee as in Definition~\ref{def:uncond-p}, but without conditional guarantees akin to Definition~\ref{def:cond-p} which makes the sequential estimate difficult to interpret. Modifying the procedure to ensure such conditional guarantees must be done through restraining or pre-setting the stopping time, which would compromise its flexibility.

\subsection{Alternative approaches}
Our discussion remains at the generality of the abstract setting described in Section~\ref{sec:setting}, where we refrain from making further assumptions on the setting. In this section, we briefly touch upon related works which showcase different approaches, which may be fruitful in specific contexts.

\begin{itemize}
\item \textbf{Exploiting knowledge on the distribution of $p^*(\bX)$.} In \cite{Fay2002, Kim2010} approaches for estimation of $\phi^*_\alpha(\bX)$ are discussed that aim to reduce the number of Monte-Carlo samples needed but require an assumption that the distribution of $p^*(\bX)$ comes from a specific class of distributions. In \cite{Silva2013} an approach to estimate $\phi^*_\alpha(\bx)$ is discussed that can bound the resampling risk with respect to the fixed-sample Monte-Carlo test as in~\eqref{eq:p_estimator_bias}, but comparisons with respect to $\phi^*_\alpha(\bx)$ are done under assumptions on the distribution of $p^*(\bX)$.
\item \textbf{Analytic approximations of the distribution of $T(\bY)$.} As an alternative to drawing random samples from $G^*(\bX)$, one may instead opt to approximate the distribution of $T(\bY)$. Naturally, this is highly dependent on the nature of the test statistic. For example, when the statistic can be expressed as a function of sample averages, the saddlepoint approximation \citep{Daniels1954} has been proven useful as an approximation tool in statistical inference \citep{Reid1988}, for example in various permutation and bootstrap-based inference \citep{Davison1988, Robinson1982, Niu2024}.
\item \textbf{Approximations of the distribution of $T(\bY)$ through limited sampling.} In \cite{Knijnenburg2009} a methodology is proposed that approximates the tails of the distribution of $T(\bY)$ through fitting a generalized Pareto distribution on a limited number of samples. Its accuracy is dependent on the true distribution of the test statistic and the value of the realized statistic $T(\bX)$.
\item \textbf{Using a smaller subgroup for tests calibrated by permutation.} For tests calibrated by permutation, the distribution $F_0(\bX)$ as in~\eqref{eq:p_val_true} is typically chosen to reflect the full invariance properties implied through the null hypothesis. However, one may opt to instead consider a smaller subgroup of permutations \citep{Koning2023, Koning2024} --- which effectively amounts to choosing a different distribution $F_0(\bX)$ within~\eqref{eq:p_val_true} --- for which the full enumeration of permutations, rather than Monte-Carlo sampling, is feasible.
\end{itemize} 

\section{Anytime-Valid Estimation of the \texorpdfstring{$p$-value}{p-value}}\label{sec:novel}
Our survey in the previous section shows nearly all existing methodology is exclusively concerned with an estimate of the testing decision $\phi^*_\alpha(\bx)$. Particularly, no existing methodology supplies a (non-granular) estimate for the $p$-value $p^*(\bX)$ which is both unconditionally valid (i.e. Definition~\ref{def:uncond-p}) as well as with prespecified $\varepsilon$-bounded ROS (i.e. Definition~\ref{def:cond-p}). Since such methodology is highly useful for practice, in this section, we aim to fill this gap by providing a proposal of our own.

Through the use of Lemma~\ref{lem:p-cond-to-uncond} it is clear that a suitable upper-confidence limit for the $p$-value can be leveraged to construct an estimate satisfying both Definition~\ref{def:uncond-p} and Definition~\ref{def:cond-p}. However, the choice of Monte-Carlo sample size remains open, and choosing the sample size that guarantees the estimate is (with high probability) not too far from the true $p$-value is problematic as it requires knowledge of $p^*(\bx)$. One could design heuristics based on multistep-approaches akin to \cite{Andrews2000, Andrews2001} discussed in Section~\ref{sec:prelim-comp}, but our goal is to find methodology for which certain validity properties can be guaranteed. 

Furthermore, we aim to develop methodology that is effective to use in practice. In light of this, we seek \emph{flexible} approaches for which sampling can be resumed at a later stage, while retaining validity guarantees. If sampling cannot be extended at a later stage, then this may be problematic for the analyst, as it requires pre-budgeting computational power for executions of the test and disallows updating estimates if more computational resources become available.

For these reasons, we propose a sequential ``anytime-valid'' methodology. Specifically, we propose a methodology which sequentially samples Monte-Carlo samples over time, providing an estimate for the $p$-value after each new sample is drawn, which can be continuously monitored and stopped for any reason at any time while retaining its validity guarantees. This also allows the methodology to be resumed at a later stage, allowing estimates to be updated after preliminary initial estimates are computed.

Our proposal is constructed through the use of \emph{confidence sequences}, which are sequential confidence intervals which retain coverage properties uniformly over time. Such sequences can be constructed in a variety of manners, and our proposed methodology is independent of the specific sequence apart from mild assumptions. We state these assumptions, and the resulting attractive properties of our methodology, in the following result:

\begin{thm}\label{th:main}
Let $B_1,B_2,\dots$ be a sequence of i.i.d. Monte-Carlo samples from $G^*(\bx)$ as in~\eqref{eq:mc_samples}. Let $\hat U_n(\bx)$ be a $\varepsilon$ upper confidence sequence for $p^*(\bx)$ based on the first $n$ samples $B_1,\dots,B_n$, i.e. a sequence such that
\[
\sup_{\bx\in\cX} \P{ \forall_n : \hat U_n(\bX) \leq p^*(\bX) \given \bX = \bx} \leq \varepsilon \ .
\]
Assume that for any fixed $\bx\in\cX$ the convergence $\hat U_n(\bx) \to p^*(\bx)$ in probability as $n\to\infty$. Now construct an estimator for $p^*(\bX)$ as:
\[
\tilde p_n(\bX) \defeq \min_{m\leq n} \hat U_m(\bX) + \varepsilon \ .
\]
For this estimator the following statements hold: 
\begin{enumerate}
\item $\tilde p_n(\bX)$ is unconditionally valid (Definition~\ref{def:uncond-p}) uniformly in $n$:
\[
\forall_{F\in\Omega_0} \forall_{t\in[0,1]} : \mathbb{P}_{\bX \sim F}\bigg( \forall_n : \tilde p_n(\bX) \leq t \bigg) \leq t \ .
\]
\item $\tilde p_n(\bx)$ has $\varepsilon$-bounded ROS (Definition~\ref{def:cond-p}) uniformly in $n$:
\[
\sup_{\bx\in\cX} \mathbb{P} \bigg( \forall_n : \tilde p_n(\bX) \leq p^*(\bX) \givenB{\bigg} \bX = \bx \bigg) \leq \varepsilon \ .
\]
\item For any $\bx\in\cX$ the estimator $\tilde p_n(\bx)$ converges in probability to the true $p$-value inflated by $\varepsilon$ as $n\to\infty$:
\[
\tilde p_n(\bx) \to p^*(\bx) + \varepsilon  \ . 
\]
\item For any $\bx\in\cX$ the estimator $\tilde p_n(\bx)$ is non-increasing:
\[
\tilde p_n(\bx) \leq \tilde p_{n-1}(\bx) \ .
\]
\end{enumerate}
\end{thm}

The proof is deferred to section~\ref{sec:proof-main}. The above theorem highlights that our proposed $p$-value estimator satisfies both characteristics for a valid $p$-value (Definition~\ref{def:uncond-p} and Definition~\ref{def:cond-p}) uniformly over the sequence of sequential samples, thereby yielding a valid $p$-value according to both definitions at any stopping time. The third property ensures that, if sufficient computational resources are available there is no persistent asymptotic power loss apart from the $\varepsilon$ value. Finally, the fourth property ensures there is no reason for the analyst to hesitate when continuing computations; newly computed $p$-value estimates cannot drift upward. In other words, if at some point a $p$-value estimate is produced which implies rejection of the null in light of the significance level set in advance by the analyst, continued samples through our procedure (thereby sharpening the $p$-value estimate) cannot lead to conflicting results on that inference. 

See Figure~\ref{fig:sketch-own} for a sketch of this methodology. The running $p$-value estimates are sketched for illustration purposes. These can be used at any time to draw inference. Two distinct runs are sketched in blue and orange. 

The above construction results in a biased $p$-value estimator. While this may initially seem undesirable, we emphasize that \emph{valid inference} is our essential main goal when computing estimates, and satisfying Definition~\ref{def:cond-p} particularly highly complicates a construction that would be unbiased. Nevertheless, as previously stated our methodology is a \emph{consistent} estimator for $p^*(\bx)+\varepsilon$.

Theorem~\ref{th:main} is stated without a particularization to any specific confidence sequence. Suitable constructions for such sequences date back to, for example, \cite{Robbins1970, Lai1976}, but sharper sequences have since been developed by, for example, \cite{WaudbySmith2022, Howard2021}. For concreteness and ease of reference, we provide a self-contained proposal of our methodology with a concrete confidence sequence in the following definition:

\begin{defi}\label{def:proposal}
Let $B_1,B_2,\dots$ be a sequence of i.i.d. Monte-Carlo samples from $G^*(\bx)$ and $S_n \defeq \sum_{i=1}^n B_i$. Let $\tilde U_n(\bx)$ be the largest solution in $p$ to:
\begin{equation}\label{eq:robbins}
\binom{n}{S_n} p^{S_n}(1-p)^{n-S_n} = \frac{\varepsilon}{n+1} \ .
\end{equation}
Then the sequence $\tilde U_n(\bX)$ is a $\varepsilon$ upper confidence sequence\footnote{This is a consequence of Ville's inequality \citep{Ville1939}, applied to a likelihood ratio of a sequence of independent binary observations \citep{Robbins1970}. The probability models in the likelihood ratio correspond either to a Bernoulli distribution with fixed parameter $p$, or uniform mixture of Bernoulli distributions over its probability parameter on $[0,1]$.} for $p^*(\bx)$ \citep{Robbins1970}, where $U_n(\bx) \to p^*(\bx)$ in probability \citep{Lai1976}. Therefore, for the construction $\tilde p_n(\bx) \defeq \min_{m=1,\dots,n}\tilde U_m(\bx) + \varepsilon$ the results in Theorem~\ref{th:main} hold.
\end{defi}

\subsection{Suggestions for stopping times}\label{sec:stoptimes}
As previously remarked, our methodology is sequential and can be stopped at any time while retaining its validity guarantees. To aid implementation we proceed by providing some useful stopping time proposals. In our context, perhaps a more suitable description would be \emph{pausing} times or \emph{reporting} times: since the methodology retains its validity guarantees regardless of the stopping time used, and may therefore be resumed afterwards --- even by a potentially different analyst. The analyst may opt to incorporate various times at which they would like to pause the procedure, and depending on the computational efforts required to reach that point decide whether further computations should be undertaken. 

\begin{itemize}
\item \emph{When inference can be drawn.} One may opt to (initially) stop the procedure when the $p$-value estimate is with high confidence on one side of the desired (and pre-specified) significance level $\alpha$. If this is the sole goal of the analyst, a methodology specifically for estimating $\phi^*_\alpha(\bx)$ may be more suitable, but since our proposed methodology may be stopped and continued at any time, software implementations could preliminarily stop at this point before prompting users whether sharper $p$-value estimates are required, in which case the estimates could be refined. A stopping time for this purpose could be
\begin{equation}\label{eq:stop_alpha}
\tau_\alpha \defeq \min\{ n : \tilde p_n(\bx) \leq \alpha , \hat L_n(\bx) > \alpha \} \ .
\end{equation}
where $\hat L_n(\bx)$ is s suitable lower confidence sequence, e.g. the smallest solution to equation~\eqref{eq:robbins} in $p$.
\item \emph{When estimates are no longer decreasing at a sufficient rate.} One may want to keep estimating until the sequential estimates of the $p$-value stop decreasing at some predetermined rate. For example
\begin{equation}\label{eq:stop_gamma}
\tau_{n_0,\gamma} \defeq \min \left\{ n : \frac{\tilde p_n(\bx) - \tilde p_{n-n_0}(\bx)}{n_0} \leq \gamma \right\} \ ,
\end{equation}
where $n_0$ and $\gamma$ can be suitably chosen by the analyst. 
\end{itemize}

For these, as well as other potential stopping times, we emphasize that these merely assist effective implementation of the methodology. The validity guarantees are retained regardless of the stopping time used, and may be optionally continued afterwards without loss of validity.

\begin{rem}
When the proposal in Definition~\ref{def:proposal} is stopped at the stopping time~\eqref{eq:stop_alpha} and only used to provide an estimate for $\phi^*_\alpha(\bx)$, the procedure particularizes to the methodology by \cite{Ding2020} if the latter methodology is adapted according to Lemma~\ref{lem:phi-cond-to-uncond}. The advantage of our proposed methodology lies in the (previously unrealized) flexibility offered; the analyst may optionally continue to estimate $p^*(\bx)$ more accurately after stopping at $\tau_\alpha$, or stopped prematurely if computations take too long, while providing a meaningful $p$-value estimate regardless of these choices.
\end{rem}

\subsection{Recommended reporting practices}
To round out our proposal, we discuss what results should be reported following the approach in Definition~\ref{def:proposal}. Denote by $\tau\in\mathbb{N}$ the stopping time chosen by the analyst. 
The analyst should always report both $\tilde p_\tau(\bx)$ and the value of $\varepsilon$ chosen. If space is available, for best practice we recommend the analyst to additionally report the value of $\tau$ and the stopping criterion used. For context, the value of a $\varepsilon$ lower confidence sequence $\hat L_n(\bx)$ can be reported as well (e.g. the smallest solution to~\eqref{eq:robbins}). 

The value of $\varepsilon$ should be chosen such that it is compatible with the maximum (asymptotic) power loss admissible for the problem, as well as the maximum admissible probability of inducing a type-I error purely through the process of Monte-Carlo calibration. A suitable default value is likely one or several orders of magnitude smaller than a target significance level $\alpha$, and for $\alpha=0.05$ we suggest $\varepsilon = 10^{-5}$. Such values of $\varepsilon$ closely aligns with the prevailing attitude of disregarding the randomness induced by the Monte-Carlo sampling when interpreting and reporting results.

While $\tilde p_\tau(\bx)$ can be interpreted without the auxiliary information as a valid $p$-value in the sense of Definition~\ref{def:uncond-p}, the value $\varepsilon$ provides a meaningful upper bound on the probability that it underestimates the true significance which helps the reader judge its reliability. Reporting $\tau$ additionally gives those seeking to sharpen the $p$-value estimate all the information needed to do so. Finally, the stopping criterion helps the reader understand to what extent it may be useful to sharpen the result; for example, if a stopping time such as $\tau_\alpha$ as in~\eqref{eq:stop_alpha} is used and $\tilde p_\tau(\bx) = 0.05$, an interested reader could justifiably wonder (and spend computational resources to investigate) if the true $p$-value is much smaller than $\alpha$, using the reported information as a starting point.

\subsection{Simulations}\label{sec:sims}
\begin{figure*}
\centering
\includegraphics[width=0.85\linewidth]{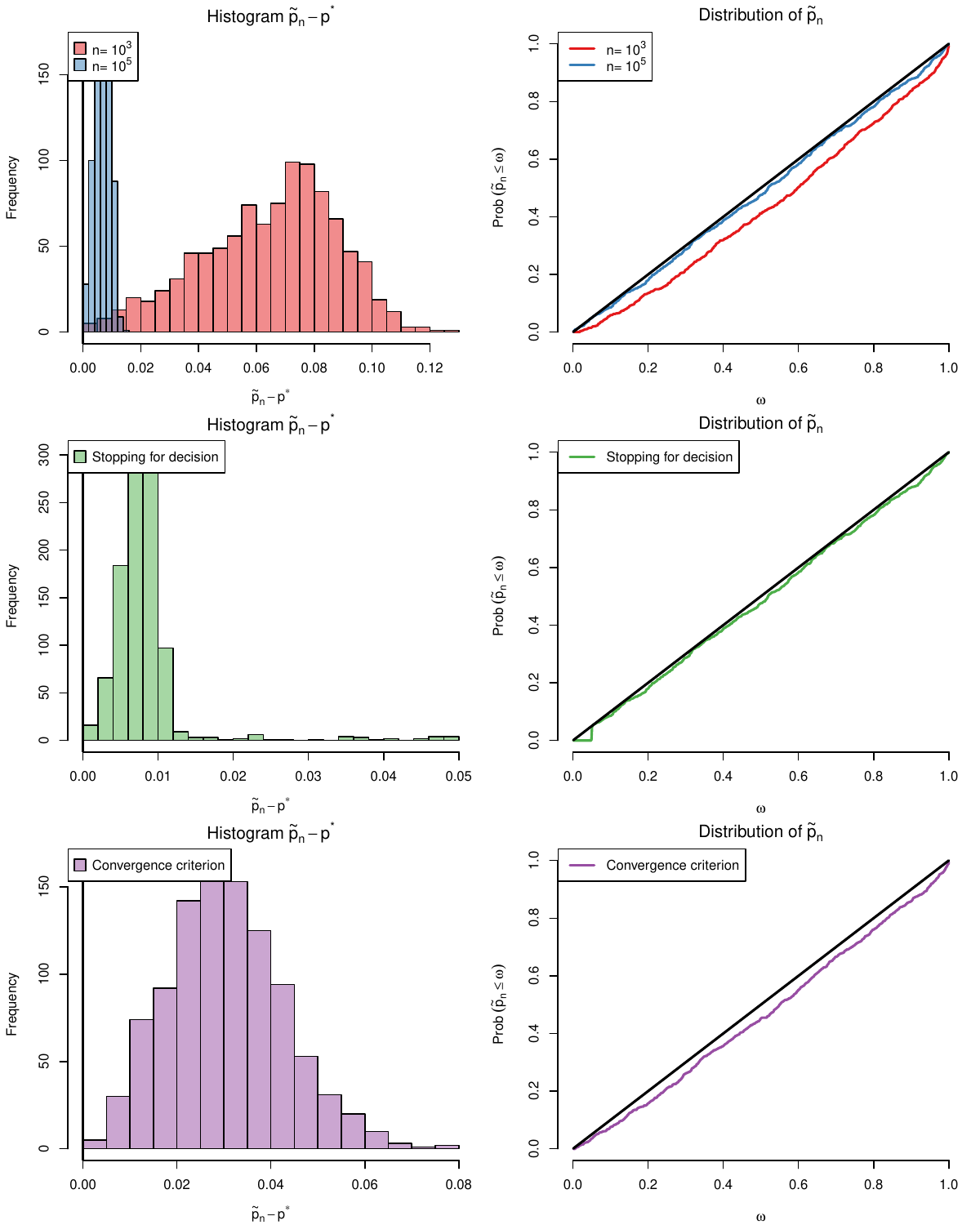}
\caption{Simulation results from the methodology in Definition~\ref{def:proposal} with $\varepsilon = 10^{-5}$ for $10^3$ independent simulation iterations. For each simulation run the true $p$-value $p^*(\bX)$ is sampled from the standard uniform distribution. With reference to the stopping times enumerated in Section~\ref{sec:sims}: the first row of plots depicts the results at predetermined stopping times \ref{en:simstop1} and~\ref{en:simstop2}; the second row the stopping time~\ref{en:simstop3}; the third row the stopping time~\ref{en:simstop4}. The plots on the left depict a histogram showing the adherence to the $\varepsilon$-bounded ROS criterion. Plots on the right show the empirical cumulative distribution function of the estimates, underscoring these stochastically dominate the uniform when $p^*(\bX)$ is uniformly distributed.} \label{fig:simulation}
\end{figure*}

In this section we showcase the appealing properties of our proposed $p$-value estimator ensuing from Definition~\ref{def:proposal} through a simulation study. In this study, we sample $p^*(\bX)$ under the null hypothesis, i.e. from a standard uniform distribution. We subsequently use the estimator from Definition~\ref{def:proposal} for a maximum of $n = 10^5$ samples and tuning $\varepsilon = 10^{-5}$. We consider the estimated value at four different stopping times:
\begin{enumerate}
\item At predetermined $n=10^3$;\label{en:simstop1}
\item At predetermined $n=10^5$;\label{en:simstop2}
\item The first time either $\tilde p_n(\bx) \leq \alpha = 0.05$ or $n=10^5$;\label{en:simstop3}
\item The first time either the criterium~\eqref{eq:stop_gamma} with $n_0=10^3$ and $\gamma = 10^{-6}$ holds, or $n=10^5$. \label{en:simstop4}
\end{enumerate}
The results are given in Figure~\ref{fig:simulation}. As can be seen, the theoretical properties given in Theorem~\ref{th:main} can be observed to hold numerically, regardless of the stopping criterion. For stopping time \ref{en:simstop1} and~\ref{en:simstop2} above, the bounds regarding conditional and unconditional validity become tighter as more samples are drawn. For stopping time~\ref{en:simstop3} above, it is reassuring to numerically observe that ``waiting out for a significant result'' cannot induce more rejections than expected under the null, and only truncates the $p$-values below $\alpha$ to the value of $\alpha$ itself. Finally, we observe how stopping time~\ref{en:simstop4} above, using the convergence criterium~\eqref{eq:stop_gamma}, results in estimates valid both conditionally and unconditionally, and may help to balance computational costs against relevant sharpening of the $p$-value estimate.

\subsection{A practical example}\label{sec:example}

In this section we discuss a practical example to showcase how our proposed methodology would embed in practice. We deliberately choose a setting where computations are not expensive (and the true $p$-value $p^*(\bx)$ can be computed) for the purposes of illustration.

We consider the data from an experiment to compare yields (measured as the dried weight of plants) obtained under a control and a treatment conditions\footnote{The dataset contains two different treatment conditions. For illustration purposes we consider only the treatment indicated as ``Trt2''.}. In total, 20 plants were used in the experiment, with half assigned to the treatment group and the other half to the control group. This dataset is discussed as an example in \cite{Dobson1983} and distributed with the statistical software R \citep{RCoreTeam2024}.

Denote the control observations by $x^{c}_1,\dots,x^{c}_{10}$ and the treatment observations by $x^{t}_1,\dots,x^{t}_{10}$. We assume these originate independently from distributions with means denoted by $\mu_c$ and $\mu_t$ respectively. We are interested to test the null hypothesis
\[
H_0: \mu_c = \mu_t \ .
\] 
Considering the test statistic $T(\bx) = \frac{1}{10}\sum_{i=1}^n x^{c}_i - x^{t}_i$. Under the above null hypothesis note that the distribution of our test statistic is exchangeable under permutations of the treatment and control labels. This suggests the significance of $T(\bX)$ can be computed as
\begin{equation}\label{eq:data-ex-p-val}
p^*(\bx) = \mathbb{P}_{\pi \sim \Pi}\bigg( T(\bX^\pi) \geq T(\bX) \givenB{\bigg} \bX=\bx \bigg) \ ,
\end{equation}
where $\Pi$ is the set of permutations of the treatment and control labels and $\pi\sim\Pi$ is sampled uniformly over $\Pi$. It is clear the setting fits into the general setting sketched in Section~\ref{sec:setting}. Instead of computing the above $p$-value exactly, in such settings one typically estimates the above value through Monte-Carlo simulation by sampling uniformly from $\Pi$, and we will do so here using the method of Definition~\ref{def:proposal}. However, given the comparatively small number of permutations $\Pi$ and the minor computational costs of computing $T(\bx)$, we are also able to compute the above $p$-value exactly through complete enumeration of $\Pi$. In fact, this is why the current setting is chosen for illustration purposes.

\begin{figure}[ht]
\centering
\includegraphics[width=\linewidth]{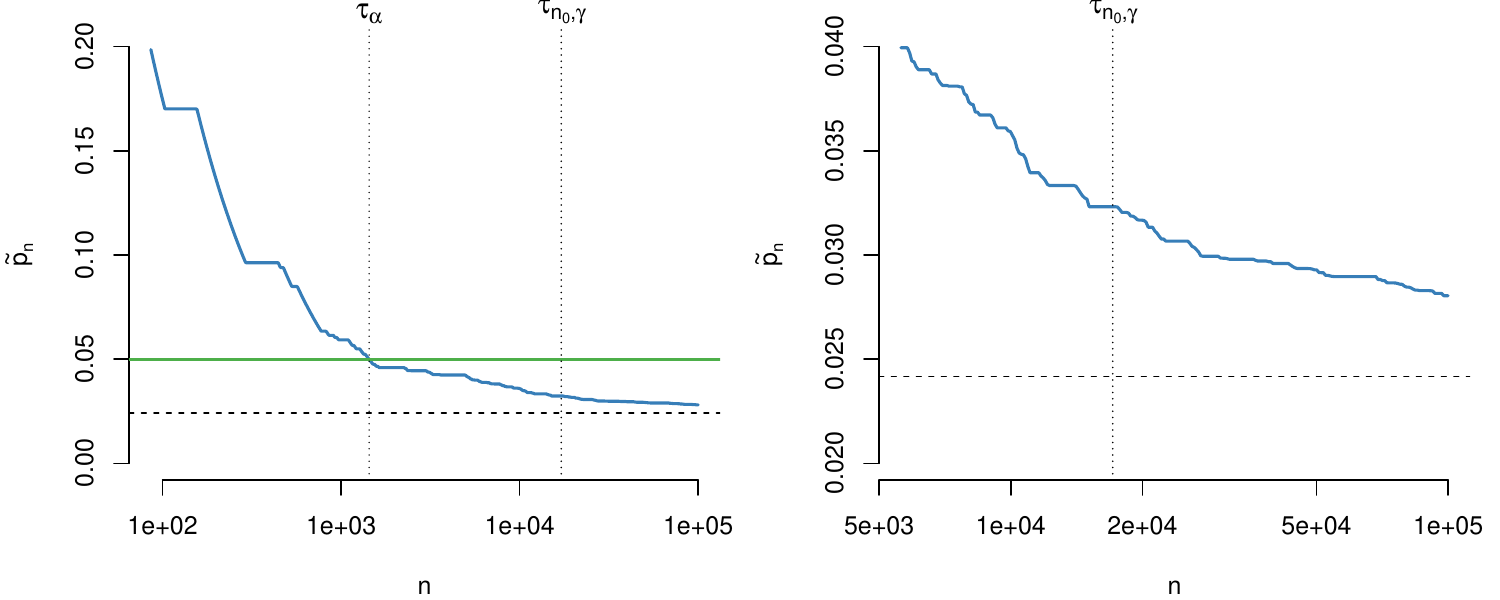}
\caption{Sequential $p$-value estimate from the methodology in Definition~\ref{def:proposal} with $\varepsilon=10^{-5}$ applied to the practical example of Section~\ref{sec:example} as a function of the sample index. The right plot is a zoomed-in version of the left plot. Vertical dotted lines correspond to the two stopping times~\eqref{eq:stop_alpha} with $\alpha=0.05$, and~\eqref{eq:stop_gamma} with $n_0 = 2\cdot 10^3$ and $\gamma = 10^{-12}$. The horizontal green line indicates the significance level $\alpha=0.05$. The horizontal dashed line corresponds to the true $p$-value $p^*(\bx) = 0.02417$ computed through complete enumeration of $\Pi$ in~\eqref{eq:data-ex-p-val}. The $x$-axes are in log-scale, and the $y$-axes are restricted to a maximum of 0.2 and 0.04 respectively. } \label{fig:realdata}
\end{figure}

The sequential estimation of the $p$-value through our methodology from Definition~\ref{def:proposal} with $\varepsilon = 10^{-5}$ is plotted in Figure~\ref{fig:realdata}, along with the two suggested stopping times from Section~\ref{sec:stoptimes}, and the exact $p$-value through complete enumeration of $\Pi$. As can be seen, the methodology converges to the true $p$-value monotonically. The first time we observe our estimate below $\alpha = 0.05$ in this run is after 1434 Monte-Carlo samples.

\begin{figure*}[ht]
\centering
\includegraphics[width=\linewidth]{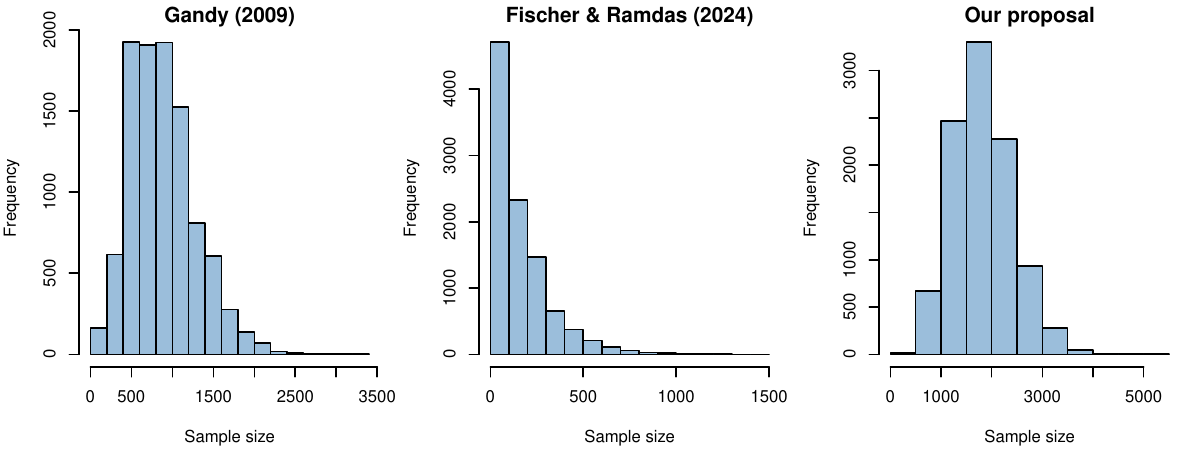}
\caption{Histograms of the Monte-Carlo sample sizes used for three methodologies applied to the practical example of Section~\ref{sec:example} for $10^4$ independent simulation runs. The method of \cite{Gandy2009} is tuned for $\alpha = 0.05$ and $\varepsilon = 10^{-3}$. The method of \cite{Fischer2024} (with reference to the original work: using the ``Binomial mixture strategy'' including stopping for futility) is tuned with $c = 0.9\alpha$ and $\alpha = 0.05$. Our methodology from Definition~\ref{def:proposal} is tuned with $\varepsilon = 10^{-5}$ and stopping time~\eqref{eq:stop_alpha} with $\alpha=0.05$.}\label{fig:sample_sizes}
\end{figure*}

One may be curious how our proposal compares to existing methodology. However, one must be careful when interpreting such comparisons, as existing methodologies have different information goals and validity guarantees. Analysts should pick methodology based on the inference goals and required validity requirements, and should \emph{not} base their decision on methodology purely on how fast a decision can be reached: such quantitative comparisons provide only limited information. Nevertheless, to give some context to the results, we repeatedly run selected approaches from related literature (tuned using parameters as the authors originally showcased them) and provide histograms of the amount of samples required before reaching a decision in Figure~\ref{fig:sample_sizes}. 

The methodology from \cite{Gandy2009} using a bounded resampling risk of $\varepsilon=10^{-3}$ stops after a mean of $885$ samples. We do not observe it failing to reject the null. This estimate comes without unconditional guarantees from Definition~\ref{def:uncond-p} (such that overall type-I error control cannot be precisely ensured) and one cannot sharpen this result further without invalidating the procedure entirely. Alternatively, the methodology from \cite{Silva2018} offers different validity guarantees. We use a set of tuning parameters (given by the authors in Table 2 in the original work) such that it has a bounded power loss of $\varepsilon = 0.005$. For this tuning we observe the methodology to require the maximum amount of samples (20 million) in $99.86\%$ of the simulation runs, terminating early (and wrongly failing to reject the null) in the remaining $0.14\%$ of the runs. Using the refined method with matching tuning parameters results in similar performance. Finally, the methodology from \cite{Fischer2024} (with reference to the original work: using the ``Binomial mixture strategy'' including stopping for futility) offers yet again different validity guarantees, and tuned with $c=0.9\alpha$, terminates after a mean of $167$ samples. It comes without any conditional guarantees in the form of Definition~\ref{def:cond-p}. In this simulation, it (wrongly) fails to reject the null in $1.72\%$ of the cases. Finally, repeatedly running our methodology results in a mean number of samples of $1821$. We do not observe it to (wrongly) fail to reject the null in any of the runs.

\vspace{-0.6cm}
\subsection{Our motivating example revisited}
We conclude the proposal of our method by showcasing how it would have fared in our motivating example of Section~\ref{sec:motex}. Specifically, recall that in this setting we found that after 30 Monte-Carlo samples had been drawn, none of these 30 samples $S_\cD(\bY_i)$ exceeded the value of the scan statistic for the original data $S_\cD(\bx)$. In other words, all samples from $G(\bX)$ in \eqref{eq:mc_samples} are zero. Since it had seemed unlikely that the final $p$-value estimate would end up larger than $\alpha=0.05$, it was tempting to stop sampling, and draw inference at that point.

It is first instructive to remark that, although 30 all-zero samples from $G(\bx)$ may feel indicative of a very small $p$-value, these 30 samples will be likely insufficient for practice. One way to see this is by considering as a $p$-value estimator the upper limit of the one-sided Clopper-Pearson confidence interval \citep{Clopper1934} based on a fixed predetermined sample. This estimator is inflexible and not recommended for practice, but its simplicity allows us to illustrate that even simple methods (which do not account for potential desired flexibility) will require much more than 30 samples unless one is willing to have very weak conditional validity guarantees. The estimator (i.e. Clopper-Pearson confidence interval) is\footnote{For clarity we present a descriptive statement of the Clopper-Pearson upper confidence limit. However, it can be computed efficiently through the relationship between the Binomial distribution and the Beta distribution \citep{Thulin2014} as
\[
\hat p_{m,\varepsilon,\text{CP}}(\bX) = F_\beta^{-1}\Big( 1-\varepsilon, \sum_{i=1}^m B_i, m- \sum_{i=1}^m B_i \Big) \ ,
\]
where $F_\beta^{-1}(q,a,b)$ is the $q$th quantile from the Beta distribution with shape parameters $a$ and $b$.
} 
\begin{equation}\label{eq:p-cp}
\hat p_{m,\varepsilon,\text{CP}}(\bX) \defeq \sup_{q \geq 0} \Bigg\{ q : \sum_{k=1}^{S_m} \binom{m}{k}q^k(1-q)^{m-k} \geq \varepsilon \Bigg\} \ ,
\end{equation}
where $S_m = \sum_{i=1}^m B_i$. Using \eqref{eq:p-cp} with $m=30$ samples, what is the minimum value of $\varepsilon$ such that this estimator could potentially result in a $p$-value estimate below $\alpha=0.05$? Clearly the estimate is lowest when all 30 samples from $G(\bX)$ are zero. In that case, the estimate is smaller than $0.05$ only if $\varepsilon > 0.215$, which is much higher than likely desirable in practice. 

Alternatively, we can ask ourselves: given \eqref{eq:p-cp} with $\varepsilon=10^{-5}$ using predetermined $m$ samples, what is the minimum amount of samples $m$ needed such that the estimator could potentially result in a $p$-value estimate below $\alpha=0.05$? In this case, the estimate is below $\alpha$ only if we observe 225 all-zero samples from $G(\bX)$. So, unless we are willing to have very weak validity guarantees we need to observe much more zeroes from $G(\bx)$ than merely 30.

With the above sample sizes for some context, we can now look at how more practicable sequential methodology would have fared. If we are only interested in the testing outcome $\phi^*_\alpha(\bx)$, the methodology from \cite{Gandy2009} is suitable. Bounding the resampling risk by $\varepsilon = 10^{-5}$, if we \emph{only} observe zeroes from $G(\bX)$, the methodology stops after 256 samples, and one can reject the null hypothesis. Compared to the fixed sample approach \eqref{eq:p-cp}, these extra samples allow the methodology flexibility to also potentially stop when a few non-zero samples from $G(\bX)$ are drawn.

After those 256 samples, one cannot further sharpen the estimate of the $p$-value, and one can only meaningfully report $\hat p(\bX) <\alpha$. If one would have liked to have the flexibility to optionally continue at this point to sharpen the $p$-value estimate, our methodology from Definition~\ref{def:proposal} should instead be used. If we again only observe zeroes from $G(\bX)$ then we would observe a $p$-value smaller than $\alpha=0.05$ after 339 samples. Compared to \cite{Gandy2009}, the extra samples needed allow for the flexibility to optionally continue to sharpen the $p$-value.

\vspace{-.3cm}
\section{Recommendations For Best Practices}\label{sec:recommend}
We summarize our findings with recommendations for best practices. For settings (in terms of combinations of inference goals and validity guarantees) where we believe analysts may be interested in, we have supplied our recommended approaches in Table~\ref{tb:recommendations}. For many settings only a single approach exists. There are two exceptions. For the outcome goal $\phi_\alpha^*(\bx)$ and unconditional validity (Definition~\ref{def:uncond-phi}), we recommend \cite{Fischer2024a} over fixed-sample size approaches based on its flexibility. For the outcome goal $\phi_\alpha^*(\bx)$ and conditional validity (Definition~\ref{def:cond-phi}), we recommend \cite{Gandy2009} over \cite{Ding2020} based on quantitative comparisons included in \cite{Ding2020}.

Analysts faced with reported results endowed with conditional validity guarantees can interpret them as follows. For the resampling risk of Definition~\ref{def:cond-phi}, the probability that an independent analyst re-running the analysis and producing a different estimate is at most $2\varepsilon$ (see Section~\ref{sec:sup-rem}), and this knowledge may even help identify potential miscomputed or excessively rare reported outcomes. For the risk of overestimated significance of Definition~\ref{def:cond-p}, if the estimate is computed correctly, then the probability that this estimate attributed too much significance to the computed statistic is bounded by $\varepsilon$. The probability that the null is rejected based solely on the Monte-Carlo sampling process is bounded by $\varepsilon$.

While we have discussed many sophisticated approaches to estimate $\phi^*_\alpha(\bX)$ or $p^*(\bX)$, with each available to bound particular types of errors of subsequent inference, we emphasize that an important source of errors is the potential (accidental) misuse of methodologies. In particular, analysts may be tempted to re-run methodologies when initial results seem off; this may, for example, occur when the analyst uses small samples and a preliminary estimate in the form of~\eqref{eq:p_estimator_bias}, which is later recomputed based on a bigger sample where previously significant results have been rendered insignificant. While there is no methodology that can prevent \emph{any} type of misuse, we can recommend the analyst to carefully consider the planned sampling strategy. In case flexibility is desirable, we recommend the analyst to consider methodology that can be stopped at any time and continued afterwards; this should prevent the need to re-run any analyses. Furthermore, estimates endowed with conditional guarantees from Definition~\ref{def:cond-phi}, Definition~\ref{def:cond-phi-power}, or Definition~\ref{def:cond-p} ensure reliability of the produced estimates which may help independent analysts interpret the outcomes.

The analyst in need to calibrate their test by Monte-Carlo simulation should choose a sampling approach based \emph{first and foremost} on the desired type of inference outcomes, and the desired validity guarantees of those outcomes. This choice should made before inspecting the data. After this choice has been settled, the analyst may be faced with multiple suitable approaches. It is \emph{only} at this stage that quantitative comparisons (for example with respect to the required amount of samples) are a meaningful tool to choose a methodology. In contrast, comparisons among methodology with varying validity guarantees are unsuitable for the analyst in need to choose a methodology and we recommend against using such comparisons for this purpose.

\begingroup
\hbadness=10000

\begin{table*}[ht]
\footnotesize
\begin{tabular}{@{}p{1.5cm}p{2.7cm}p{3.1cm}p{3.9cm}p{3.5cm}@{}}
\toprule
\textbf{Goal} &
  \textbf{Validity} &
  \textbf{Recommendation} &
  \textbf{Comments} & \textbf{Reporting}\\ \midrule
$\phi^*_\alpha(\bx)$ &
  Unconditional (Def. \ref{def:uncond-phi}) &
  \cite{Fischer2024} &
  Can be stopped at any time while retaining validity. & $\hat \phi_\alpha(\bx)$ and the number of samples used. Do not report any estimate for $p^*(\bx)$.\\
 &
  Conditional (Def. \ref{def:cond-phi}, see also Lemma~\ref{lem:rr-implies-near-uncond}) &
  \cite{Gandy2009} &
  Cannot be stopped prematurely and may require a long runtime if $p^*(\bx) - \alpha$ is small.  & $\hat \phi_\alpha(\bx)$ and $\varepsilon$. Do not report any estimate for $p^*(\bx)$. \\
   &
  Unconditional and conditional bounded power loss (Def. \ref{def:uncond-phi} and Def. \ref{def:cond-phi-power}) &
  \cite{Silva2018} &
  Default tuning parameters may lead to a substantial amount of simulations. Consider computational overhead before using. & $\hat \phi_\alpha(\bx)$ and $\varepsilon$. If space is available, report all tuning parameters. Do not report any estimate for $p^*(\bx)$. \\
\midrule
$p^*(\bx)$ at prespecified granularity &
  Conditional \newline(Def. \ref{def:cond-p}) &
  \cite{Gandy2020} & Cannot be stopped prematurely. If this flexibility is needed, consider method from Definition~\ref{def:proposal} which additionally requires no prespecified granularity and comes with conditional validity (Definition~\ref{def:cond-p}) but may require more samples. & $\hat p(\bx)$ as the upper end of the bucket selected, and $\varepsilon$. If space is available, report (for context only) the lower end of the bucket selected.
   \\
 &
 Unconditional and conditional (Def. \ref{def:uncond-p} and \newline Def. \ref{def:cond-p}) & 
  \cite{Gandy2020} modified through Lemma~\ref{lem:p-cond-to-uncond} & Cannot be stopped prematurely. If this flexibility is needed, consider method from Definition~\ref{def:proposal} which additionally requires no prespecified granularity but may require more samples. & $\hat p(\bx)$ as the upper end of the bucket increased by $\varepsilon$, and the value of $\varepsilon$ separately. If space is available, report (for context only) the lower end of the bucket selected.
   \\ \midrule
$p^*(\bx)$ & Unconditional and conditional \newline (Def. \ref{def:uncond-p} and \newline Def. \ref{def:cond-p}) & Our proposal in Definition~\ref{def:proposal}                    & Can be stopped at any time while retaining validity. & $\hat p_\tau(\bx)$ and $\varepsilon$. If space is available, report criterion used for $\tau$, its value, and (for context only) the value of the lower end of a confidence sequence $\hat L_n(\bx)$ at level $\varepsilon$ (e.g. smallest solution to~\eqref{eq:robbins}). \\ \bottomrule
\end{tabular}\caption{Recommendations for combinations of inference goals and validity guarantees.}\label{tb:recommendations}
\end{table*}

\endgroup

\FloatBarrier
\newpage
\appendix
\section{Appendix}
\subsection{Remark on Resampling Risk}\label{sec:sup-rem}
The reader is advised to carefully interpret the word ``resampling'' when referring to resampling risk. Related literature sometimes discusses the resampling risk in correspondance with ``Gleser's first law of applied statistics'' stating that ``Two individuals using the same statistical method on the same data should arrive at the same conclusion'' (Comment to \cite{DiCiccio1996}). However, bounding this risk does not directly bound the \emph{replicability} between analysts. Rather it bounds the probability of disagreeing with the actual (unknown) test outcome $\phi_\alpha^*(\bx)$. Nevertheless, these are intimately related events. Subsequently, we can bound the probability that two separate analysts with the same observations $\bx$, test statistic and significance level, producing estimates for the test outcome $\hat \phi_\alpha^{(1)}(\bx)$ and $\hat \phi_\alpha^{(2)}(\bx)$, with resampling risk bounded by $\varepsilon$, will disagree based (purely) on the randomness due to Monte-Carlo calibration as follows:
\begin{align*}
\P{\hat \phi_\alpha^{(1)}(\bx) \neq \hat\phi_\alpha^{(2)}(\bx) \given \bX = \bx } &= \P{\hat \phi_\alpha^{(1)}(\bx) = \phi^*_\alpha(\bx), \hat\phi_\alpha^{(2)}(\bx) = 1-\phi^*_\alpha(\bx) \given \bX = \bx} \\
&\hspace{0.5cm} + \P{\hat \phi_\alpha^{(1)}(\bx) = 1-\phi^*_\alpha(\bx), \hat\phi_\alpha^{(2)}(\bx) = \phi^*_\alpha(\bx) \given \bX = \bx} \\
& \leq \varepsilon + \varepsilon = 2\varepsilon \ .
\end{align*}

\subsection{Discussion on connection between unconditional validity and conditional validity}\label{sec:connecting-validities-phi}
As noted in Section~\ref{sec:validity-phi}, following the discussion of Lemma~\ref{lem:rr-type1-bounds}, one may wonder if there is a possibility to construct an estimator, based on an estimator with some bounded resampling risk, such that the newly constructed estimator has both a bounded resampling risk as well as type-I error control. Specifically, a construction of the form $\tilde\phi_\alpha := \hat\phi_\alpha$ where $\hat\phi_\alpha$ has some some well-chosen uniformly bounded resampling risk (possibly dependent on both $\varepsilon$ and $\alpha$) such that $\tilde\phi_\alpha$ has \emph{both} a $\varepsilon$ bounded resampling risk and overall type-I error bounded by $\alpha$. The following lemma shows that this is impossible without further assumptions on the test $\hat\phi_\alpha$ which go beyond the (uniform) bound of the resampling risk:

\begin{lem}\label{lem:rr-type1-bounds}
Let $\hat\phi_\alpha(\bX)$ be an estimator for the testing outcome $\phi^*_\alpha(\bX)$ with $\varepsilon$-bounded resampling risk as in Definition~\ref{def:cond-phi}. Define the set of observations:
\[
\cX_\alpha \defeq \{ \bx\in\cX : \phi^*_\alpha(\bx) = 1 \} \ .
\]
Then for all $F\in\Omega_0$ the overall type-I error of this estimator can be lower bounded by
\begin{align*}
\mathbb{P}_{\bX \sim F}\left( \hat \phi_{\alpha}(\bX) = 1 \right) &\geq (1-\varepsilon)\mathbb{P}_{\bX\sim F}\Big(\phi^*_\alpha(\bX) = 1\Big) \\ &\hspace{0.5cm} + \int_{\bx \not\in \cX_\alpha} \P{\hat\phi_\alpha(\bX) = 1 \given \bX = \bx} {\rm d}F(\bx) \ , \numberthis\label{eq:rr-type1-lower}
\end{align*}
and upper bounded by
\begin{equation}\label{eq:rr-type1-upper}
\mathbb{P}_{\bX \sim F}\left( \hat \phi_{\alpha}(\bX) = 1 \right) \leq \varepsilon + \int_{\bx \in \cX_\alpha} \mathbb{P}\Big(\hat\phi_\alpha(\bX) = 1 \givenB{\Big} \bX = \bx\Big) {\rm d}F(\bx) \ .
\end{equation}
\end{lem}

The proof is deferred to the end of this section. Depending on the exactness of the true test $\phi^*_\alpha(\bX)$, as well as the probability of $\hat\phi_\alpha(\bx)$ estimating a rejection when $\bx\not\in\cX_\alpha$ the above lower bound may preclude unconditional validity for the test. For example, if the true test $\phi^*_\alpha(\bx)$ is an exact test, i.e.
\[
\mathbb{P}_{\bX\sim F}\Big(\phi^*_\alpha(\bX) = 1\Big) = \alpha \ ,
\] 
and the probability of rejection of the estimator $\hat\phi_\alpha(\bx)$ for $\bx\not\in\cX_\alpha$ is equal to the maximum value admissible under the assumption that it has $\varepsilon$-bounded resampling risk, i.e. for all $\bx\in\cX_\alpha$:
\[
\mathbb{P}\Big(\hat\phi_\alpha(\bX) = 1 \givenB{\Big} \bX = \bx\Big) = \varepsilon \ ,
\]
then the lower bound~\eqref{eq:rr-type1-lower} is equal to 
\[
(1-\varepsilon)\alpha + (1-\alpha)\varepsilon = \alpha + \varepsilon(1-2\alpha) \ .
\]
This lower bound exceeds $\alpha$ whenever $\alpha \leq 1/2$ irrespective of the choice of $\varepsilon > 0$. In other words, for this test (which has $\varepsilon$-bounded resampling risk) unconditional validity is impossible regardless of the choice of $\varepsilon > 0$.

The lemma above precludes any hope of finding a construction of the form $\tilde\phi_\alpha := \hat\phi_\alpha$ where $\hat\phi_\alpha$ has some some well-chosen uniformly bounded resampling risk (possibly dependent on both $\varepsilon$ and $\alpha$) such that $\tilde\phi_\alpha$ has both a $\varepsilon$ bounded resampling risk and overall type-I error bounded by $\alpha$: such a result necessarily requires knowledge of an upper bound on $\mathbb{P}(\hat\phi_\alpha(\bX) = 1 \givenB{} \bX = \bx)$ depending on whether $\bx\in\cX_\alpha$ or not. This is more fine-grained knowledge, which goes beyond the \emph{uniform} $\varepsilon$-bounded resampling risk.

\textit{Proof of Lemma~\ref{lem:rr-type1-bounds}.}
Let $F\in\Omega_0$ arbitrary but fixed. Then we can expand our probability of interest as:
\begin{align*}
\mathbb{P}_{\bX\sim F}\Big(\hat\phi_\alpha(\bX)= 1 \Big) &= \int_{\bx\in\cX_{\alpha}}\mathbb{P}\Big(\hat\phi_\alpha(\bX)= 1 \givenB{\Big} \bX = \bx\Big){\rm d}F(\bx) \\
&+ \int_{\bx\not\in\cX_{\alpha}}\mathbb{P}\Big(\hat\phi_\alpha(\bX)= 1  \givenB{\Big} \bX = \bx\Big){\rm d}F(\bx) \ . \numberthis\label{eq:type1-expanded}
\end{align*}
For the upper bound result, note that for all $\bx\not\in\cX_\alpha$ the $\varepsilon$-bounded resampling risk implies 
\[
\mathbb{P}\Big(\hat\phi_\alpha(\bX)= 1  \givenB{\Big} \bX = \bx\Big) \leq \varepsilon  \ ,
\]
which implies the bound
\[
\int_{\bx\not\in\cX_{\alpha}}\mathbb{P}\Big(\hat\phi_\alpha(\bX)= 1  \givenB{\Big} \bX = \bx\Big){\rm d}F(\bx) \leq \varepsilon \ .
\]
Using the above bound in~\eqref{eq:type1-expanded} proves the upper bound result~\eqref{eq:rr-type1-upper} of the lemma. For the lower bound, note that the $\varepsilon$-bounded resampling risk implies for all $\bx\in\cX_\alpha$:
\[
\mathbb{P}\Big(\hat\phi_\alpha(\bX)= 0  \givenB{\Big} \bX = \bx\Big) \leq \varepsilon \ .
\]
The above result can be used to lower bound the first component in~\eqref{eq:type1-expanded} as follows:
\[
\int_{\bx\in\cX_{\alpha}}\mathbb{P}\Big(\hat\phi_\alpha(\bX)= 1  \givenB{\Big} \bX = \bx\Big){\rm d}F(\bx) \geq (1-\varepsilon)\int_{\bx\in\cX_{\alpha}} {\rm d}F(\bx) \ .
\]
Substituting the above bound in~\eqref{eq:type1-expanded} implies the lower bound result~\eqref{eq:rr-type1-lower} of the lemma.
\hfill $\square$

\subsection{Proof of Lemma~\ref{lem:rr-implies-near-uncond}}
\begin{proof}
The proof is a simple consequence of the upper bound~\eqref{eq:rr-type1-upper} of Lemma~\ref{lem:rr-type1-bounds}, by observing that if $F\in\Omega_0$:
\[
\int_{\bx \in \cX_\alpha} \mathbb{P}(\hat\phi_\alpha(\bX) = 1 \givenB{} \bX = \bx) {\rm d}F(\bx) \\
\leq \int_{\bx \in \cX_\alpha} 1\cdot {\rm d}F(\bx) \leq \alpha \ ,
\]
where the second inequality follows from $F\in\Omega_0$. Combining the above bound with~\eqref{eq:rr-type1-upper} implies the result.
\end{proof}

\subsection{Proof of Lemma~\ref{lem:p-cond-to-uncond}}\label{sec:p-cond-to-uncond}
\begin{proof}
Let $F\in\Omega_0$ and $t\in[0,1]$ arbitrary and fixed. First we define a subset of $\cX$ as
\begin{equation}\label{eq:subset-suppX}
\cX_{t-\varepsilon} \defeq \{ \bx\in\cX : p^*(\bx) \leq t-\varepsilon \} \ .
\end{equation}
Then observe that, if $\bx\not\in\cX_{t-\varepsilon}$ for the estimator $\hat p(\bX)$ we have:
\begin{align*}
\P{ \hat p(\bX) \leq t-\varepsilon \given \bX = \bx} \leq \P{ \hat p(\bX) \leq p^*(\bx) \given \bX = \bx} \leq \varepsilon\ ,
\end{align*}
where the second inequality follows from the bounded ROS assumption. We can complete the proof by expanding our probability of interest as follows:
\begin{align*}
\mathbb{P}_{\bX\sim F}\Big(\tilde p(\bX) \leq t \Big) &= \mathbb{P}_{\bX\sim F}\Big(\hat p(\bX) \leq t - \varepsilon \Big) \\
&= \int_{\bx\not\in\cX_{t-\varepsilon}}\mathbb{P}\Big(\hat p(\bX) \leq t - \varepsilon \givenB{\Big} \bX = \bx\Big){\rm d}F(\bx) \\
&+ \int_{\bx\in\cX_{t-\varepsilon}}\mathbb{P}\Big(\hat p(\bX) \leq t - \varepsilon \givenB{\Big} \bX = \bx\Big){\rm d}F(\bx) \\
&\leq \varepsilon + \int_{\bx\in\cX_{t-\varepsilon}}{\rm d}F(\bx) \ . \\
&= \varepsilon + \mathbb{P}_{\bX\sim F}\Big( p^*(\bX) \leq t-\varepsilon \Big) \\
&\leq \varepsilon + (t-\varepsilon) = t \ ,
\end{align*}
where the final inequality follows since $F\in\Omega_0$.
\end{proof}

\subsection{Proof of Theorem~\ref{th:main}}\label{sec:proof-main}
\begin{proof}
Statement 3 follows directly by the convergence assumption on $\hat U_n(\bx)$. Statement 4 trivially follows from the monotonicity of the running minimum.

What is left to prove are the first two statements. A crucial remark is that the running minimum of $\hat U_{n}(\bx)$ is also a $\varepsilon$ upper confidence sequence for  $p^*(\bx)$, i.e.
\begin{equation}\label{eq:runmin-confseq}
\sup_{\bx\in\cX} \P{ \forall_n : \min_{m\leq n} \hat U_m(\bx) \leq p^*(\bx) \given \bX = \bx} \leq \varepsilon \ .
\end{equation}
This fact is frequently stated in related literature \citep{Darling1967}, but we include an explicit proof here for clarity. Define (conditional on $\bX=\bx$) the event $A_n := \{ \hat U_n(\bx) < p^*(\bx) \}$ and note that if $\hat U_n(\bx)$ is a $\varepsilon$-upper confidence sequence this implies $\P{ \bigcap_{n=1}^\infty A_n } \leq \varepsilon$.
Now note that~\eqref{eq:runmin-confseq} can equivalently be written as
\[
\sup_{\bx\in\cX}\P{ \bigcap_{n=1}^\infty \bigcap_{m=1}^n A_m } \leq \varepsilon \ .
\] 
Since $\bigcap_{n=1}^\infty \bigcap_{m=1}^n A_m = \bigcap_{n=1}^\infty A_n$, the statement of~\eqref{eq:runmin-confseq} follows. Ultimately, since $\varepsilon > 0$, we have $\tilde p_n (\bx) \geq \min_{m\leq n} \hat U_m(\bx) $ and this directly implies statement 2. 

The proof of statement 1 is analogous to the arguments of Lemma~\ref{lem:p-cond-to-uncond}. For brevity, define $\bar U_n(\bx) := \min_{m\leq n} \hat U_m(\bX)$. Letting $F\in\Omega_0$ and $t\in[0,1]$ be arbitrary and recalling the definition in~\eqref{eq:subset-suppX} we have:
\begin{align*}
 \mathbb{P}_{\bX\sim F}\Big(\forall_n : \tilde p_n(\bX) \leq t \Big) &= \mathbb{P}_{\bX\sim F}\Big(\forall_n : \bar U_n(\bx) \leq t - \varepsilon \Big) \\
&= \int_{\bx\not\in\cX_{t-\varepsilon}}\mathbb{P}\Big(\forall_n : \bar U_n(\bX) \leq t - \varepsilon \givenB{\Big} \bX = \bx\Big){\rm d}F(\bx) \\
&+ \int_{\bx\in\cX_{t-\varepsilon}}\mathbb{P}\Big(\forall_n : \bar U_n(\bX) \leq t - \varepsilon \givenB{\Big} \bX = \bx\Big){\rm d}F(\bx) \\
&\leq \varepsilon + \mathbb{P}_{\bX\sim F}\Big( p^*(\bX) \leq t-\varepsilon \Big) \\
&\leq \varepsilon + (t-\varepsilon) = t \ ,
\end{align*}
where we used~\eqref{eq:runmin-confseq} in the first inequality, implying statement 1, and completing the proof.
\end{proof}

\section*{Acknowledgements}
We would like to thank Muriel F. Pérez-Ortiz for carefully reading an earlier version of the manuscript and for useful discussions.

\FloatBarrier
\phantomsection\addcontentsline{toc}{section}{References}
\small
\urlstyle{rm}
\interlinepenalty=10000
\bibliographystyle{custom_bib}
\bibliography{ep_bib}

\end{document}